\documentstyle[11pt,fklamb-pasp,twoside,epsf]{article} 
\begin{document}
\title{High Frequency QPOs in Neutron Stars and Black Holes: Probing Dense Matter and Strong Gravitational Fields}
 \author{Frederick K. Lamb}
\affil{Center for Theoretical Astrophysics, Department of Physics, and Department of Astronomy, University of Illinois at Urbana-Champaign,\\ 1110 W. Green Street, Urbana, IL 61801-3080, USA; fkl@uiuc.edu}

\begin{abstract}
Quasi-periodic oscillations (QPOs) have been discovered in the X-ray emission of many neutron stars and black holes. The QPOs with frequencies $\ga 300$~Hz are thought to be produced near the surfaces of neutron stars and the event horizons of black holes. I first summarize some of the most important properties of the QPOs seen in neutron star and black hole systems. I then review some of the models that have been proposed and compare them with observational data. Finally, I describe how these QPOs can be used to determine the properties of dense matter and strong gravitational fields.
\end{abstract}

\section{Introduction}

It is an honor to be able to contribute to this volume memorializing Jan van Paradijs. Jan made important scientific contributions to many areas of astrophysics, but especially to our understanding of thermonuclear X-ray bursts and the properties of low-mass X-ray binary systems (LMXBs). It is therefore particularly fitting to include a few words about Jan's scientific contributions in this review, which is concerned with X-ray oscillations in LMXBs.

Jan was one of the first to recognize that careful study of X-ray bursts could provide quantitative information about the properties of neutron star matter and strong gravitational fields. He realized that the {\em Rossi X-Ray Timing Explorer} ({\em RXTE\/}) would make possible much more sensitive and detailed studies of these questions. Jan led a collaboration that anticipated the detection of periodic brightness oscillations during thermonuclear X-ray bursts and 10--60~Hz QPOs in the atoll sources. Even before {\em RXTE\/} was launched, he and his collaborators requested and were awarded observing time to search for these two phenomena.  Both were discovered soon after {\em RXTE\/} was launched and are among the most important discoveries made with it. Jan's team was somewhat unlucky in not catching any bursts with detectable oscillations early in the mission, but they discovered kilohertz QPOs in many neutron star X-ray sources. As I describe below, these and other discoveries made with {\em RXTE\/} have opened a new era in the study of dense matter and strong gravitational fields. I continue to miss Jan as a friend as well as a scientific colleague.

Neutron stars and black holes are important cosmic laboratories for studying fundamental questions in physics and astronomy, especially the properties of dense matter and strong gravitational fields. The discovery using {\em RXTE\/} of oscillations in the X-ray emission of accreting neutron stars and black holes with frequencies comparable to the dynamical frequencies near these objects has provided important new tools for studying these questions.

The neutron stars in LMXBs have proved to be particularly valuable systems for investigating the innermost parts of accretion disks, gas dynamics and radiation transport in strong radiation and gravitational fields, and the properties of dense matter, because the magnetic fields of many of these stars are relatively weak ($\sim\,$10$^7$--$10^{10}$~G) but not negligible. Magnetic fields of this size are weak enough to allow at least a fraction of the accreting gas to remain in orbit as it moves into the strong gravitational and radiation fields of these stars, but strong enough to produce anisotropic X-ray emission, potentially allowing the spin rates of these stars to be determined. The ``clockwork'' of these systems is remarkably regular and systematic. Observations using {\em RXTE\/} are providing a wealth of diagnostic information about strong gravitational fields and the magnetic fields, spin rates, masses, and radii of these stars (for previous reviews, see Lamb, Miller, \& Psaltis 1998a, 1998b; van der Klis 1998, 2000; Psaltis 2000). 

QPOs have also been discovered in the X-ray emission of accreting black holes in LMXBs. Study of this variability is providing increasingly detailed information about these objects. The recent discovery that \hbox{GRO~J1655$-$40} and \hbox{GRS~1915$+$105} both have two high-frequency QPOs (Remillard et al.\ 2001; Strohmayer 2001c, 2001d) is the first evidence that high-frequency QPOs also occur in pairs in black hole systems and the strongest evidence to date that the black hole \hbox{GRO~J1655$-$40} has significant angular momentum.

The discovery with {\em RXTE\/} of X-ray variability on dynamical time scales in black holes and neutron stars has revolutionized the study of accretion and X-ray emission by these objects. Before the launch of {\em RXTE}, Newtonian models were often used; after the discovery of very high frequency oscillations with {\em RXTE}, strong-field general relativistic calculations have become the norm, to allow quantitative comparisons of models with the {\em RXTE\/} data. The highest-frequency QPOs are thought to come from regions where the gravitational field is very strong. Study of these oscillations may provide the best near-term opportunity to observe and measure uniquely general relativistic effects, such as gravitomagnetic\footnote{The term ``gravitomagnetic'' refers to the close analogy between the motion of a test particle in the spacetime near a circulating mass current and the motion of a charged particle in the magnetic field near a circulating electrical current. No magnetic field is involved.} (Lense-Thirring) precession and the absence of stable circular orbits around sufficiently compact relativistic objects, and to determine the properties of the dense matter in neutron stars. {\em RXTE\/} continues to provide a wealth of new information on conditions close to the surfaces of neutron stars and the event horizons of black holes and to stimulate imaginative new thinking about the physical processes that dominate in these extreme conditions.

In \S2 I summarize some of the most important properties of the X-ray oscillations seen in neutron star and black hole systems. In \S3 I review some of the mechanisms that have been proposed for generating these oscillations, compare them with some of the data on neutron star and black hole QPOs, and discuss the similarities and differences of QPO generation in neutron star and black hole systems. In \S4 I describe how measurements of the highest-frequency QPOs can be used to determine the properties of dense matter and strong gravitational fields. In \S5 I conclude with a few remarks about the progress we have made in understanding black hole and neutron star QPOs and using them as tools to better understand the properties of strong gravitational fields and dense matter. There is space here to review only a fraction of the large body of work on these problems. I apologize to those whose work is not adequately discussed.

\section {Properties of Neutron Star and Black Hole QPOs}
\label{sec:properties}

This section summarizes several important properties of the oscillations that have been detected in the X-ray emission of accreting neutron stars and black holes. More detailed reviews of the observations may be found in van der Klis (1998, 2000).

\subsection{Brightness Oscillations in Neutron Stars}
\label{sec:NS-Osc}

Three classes of X-ray brightness oscillations have been detected in neutron stars in LMXBs: nearly periodic oscillations that, with one exception, have been detected only during thermonuclear X-ray bursts; pairs of kilohertz QPOs in accretion-powered emission; and several lower-frequency QPOs, also in accretion-powered emission. A key feature of the QPOs and peaked noise components seen in neutron stars is that the frequencies of all of them (except the recently identified hectohertz peaked noise component) vary significantly and are positively correlated with each other and with the inferred X-ray flux (see, e.g., van Straaten et al.\ 2001).

{\em X-ray burst oscillations}.---X-ray flux oscillations with frequencies ranging from $\sim\,$330~Hz to $\sim\,$590~Hz have been detected during some of the thermonuclear X-ray bursts of ten neutron stars (see Strohmayer 2001a, 2001b). These oscillations remain nearly sinusoidal and relatively stable ($\Delta\nu/\nu \sim 0.01$) even when their peak-to-peak amplitudes are $\sim\,$70\% and the luminosity of the star approaches the Eddington critical luminosity at which the outward force of the radiation balances the inward force of gravity. This indicates that they are not caused by oscillations of the stellar surface or the accretion disk, because such oscillations would become nonlinear at high amplitudes and would also be affected by strong radiation forces. Their stability is easier to understand if their frequencies are related to the spin of the star, which is very stable.

The X-ray spectra, waveforms, and long-term stability of these oscillations indicate that they are generated by rotation of a brighter region, or two antipodal brighter regions, of the stellar surface (Strohmayer et al.\ 1996, 1997, 1998; Bildsten 1998; Miller \& Lamb 1998; Markwardt, Strohmayer, \& Swank 1999; Miller 1999; Strohmayer \& Markwardt 1999; Strohmayer 1999, 2001a, 2001b; Weinberg, Miller, \& Lamb 2001). If so, the frequency $\nu_{\rm osc}$ of the strongest oscillation is close to the stellar spin frequency $\nu_{s}$ or its second harmonic. The frequencies of the burst oscillations are comparable to the spin frequencies predicted by the recycling hypothesis for millisecond rotation-powered pulsars and the theory of disk accretion by weakly magnetic neutron stars (see Ghosh \& Lamb 1992 and references therein). 

Several properties of the burst oscillations are not yet understood. These properties include the relatively large fractional changes in frequency (up to $\sim\,$1.3\%) seen in some sources (Galloway et al. 2000), the simultaneous presence of several distinct frequencies (Miller 2000a; Galloway et al. 2000; Chakrabarty 2000), phase shifts during some bursts that are $\sim\,$10$\pi$ (Miller 1999b), and very low upper limits on the amplitudes of higher harmonics of the basic oscillation frequency in some sources (Strohmayer 2001a, 2001b).

A strong (amplitude $\sim\,$50\%) brightness oscillation has recently been found in data taken during a 1997 thermonuclear X-ray burst from \hbox{SAX~J1808$-$369} (in't Zand et al.\ 2001). This source is an accretion-powered X-ray pulsar (Wijnands \& van der Klis 1998) in a low-mass binary system (Chakrabarty \& Morgan 1998). The frequency of the burst oscillation is within a few Hertz of the 401~Hz spin frequency inferred from the periodic modulation of its accretion-powered emission. The burst oscillation could be produced by a mechanism different from the mechanism that produces the burst oscillations seen in other neutron-star LMXBs, but this seems unlikely. Thus, although the detailed physics of the burst oscillations is not yet understood, the evidence that they are generated by the spin of the neutron star is strong.

{\it Kilohertz QPOs}.---QPOs with frequencies between $\sim\,$500~Hz and $\sim\,$1300~Hz have now been detected in the accretion-powered emission of more than twenty neutron stars in LMXBs (see van der Klis 2000). They usually appear as a pair of prominent, narrow peaks in power spectra. The frequencies $\nu_1$ and $\nu_2$ of the lower and upper kilohertz QPOs move upward and downward together and can increase and decrease by as much as a factor of two within a few hundred seconds while remaining highly coherent ($\nu/{\rm FWHM} \sim 100$; see van der Klis 1995 and Wijnands et al.\ 1998), indicating that the QPO mechanism is highly tunable. Their frequencies are strongly positively correlated with the X-ray flux. In some sources, the frequency separation ${\Delta\nu} \equiv \nu_2 - \nu_1$ between the two kilohertz QPOs decreases systematically by 30--100~Hz, depending on the source, as $\nu_2$ increases by a much larger amount (\hbox{Sco~X-1}: van der Klis et al.\ 1997; \hbox{4U~1608$-$52}: M\'endez et al.\ 1998; \hbox{4U~1735$-$44}: Ford et al.\ 1998; \hbox{4U~1728$-$34}: M\'endez \& van der Klis 1999; see also Psaltis et al.\ 1998b). The kilohertz QPOs have rms amplitudes in the 2--60~keV band of the {\em RXTE\/} Proportional Counter Array ranging from $\la 1$\% up to $\sim\,$15\%  and quality factors $Q \equiv \nu/{\rm FWHM}$ as large as $\sim\,$200. Sidebands of the lower kilohertz QPO have been detected in three sources (Jonker, M\'endez, \& van der Klis 2000).

\begin{table}[t]
\begin{center}
\begin{minipage}{130mm}
\caption{Frequencies of Burst Oscillations and Kilohertz QPOs$^a$}\label{table.BurstOsc}
\vspace{3pt}
\begin{tabular}{ccccc}
\tableline
\tableline
\noalign{\kern 4pt}
Source   &$\nu_{\rm osc, max}$~(Hz)\ \  &\ \ ${\Delta\nu}_{\rm max}$~(Hz)\ \   &\ $\nu_{\rm osc}/{\Delta\nu}_{\rm max}$\ \ \ \  &References \\
\noalign{\kern 3pt}
\tableline
\noalign{\kern 3pt}
\hbox{4U~1702$-$43}    &$330.55 \pm 0.02$    &$344 \pm 7$  &$0.96 \pm 0.02$ &1\\
\noalign{\kern 1pt}
\hbox{4U~1728$-$34}    &$364.23 \pm 0.05$    &$349.3 \pm 1.7$  &$1.043 \pm 0.005$ &2\\
\noalign{\kern 1pt}
\hbox{4U~1608$-$52}    &619     &$301.3 \pm 7.9$   &$2.05 \pm 0.013$ &3,4,5\\
\noalign{\kern 1pt}
\hbox{4U~1636$-$53}   &$581.75 \pm 0.13$  &$254 \pm 5$  &$2.29 \pm 0.04$ &6\\
&290$^b$    &&1.14 &7\\
\noalign{\kern 1pt}
\hbox{KS~1731$-$26}    &$523.92 \pm 0.05$     &$260 \pm 10$  &$2.015 \pm 0.077$  &8\\
\noalign{\kern 4pt}
\tableline
\noalign{\kern 4pt}
\end{tabular}
$^a$Updated from van der Klis (2000). $^b$The oscillation at 290~Hz appears to be intermittent (see Strohmayer 2001a, 2001b); when present, its amplitude is $\sim\,$40\% of the amplitude of the oscillation at 580~Hz. References: [1]~Markwardt et al.\ (1999) [2]~M\'endez \& van der Klis (1999) [3]~Chakrabarty et al.\ (2000) [4]~M\'endez et al.\ (1997) [5]~M\'endez (2000, personal communication) [6]~M\'endez et al.\ (1998) [7]~Miller (1999) [8]~Wijnands \& van der Klis (1997)
\end{minipage}
\end{center}
\end{table}

Kilohertz QPOs have been detected with high confidence in five sources that also produce burst oscillations (see Table~\ref{table.BurstOsc}). In all five, the burst oscillation frequency $\nu_{\rm osc}$ is within 4\% of the maximum observed value of ${\Delta\nu}$ or within 0.35\%--15\% of $2{\Delta\nu}_{\rm max}$. There are no sources in which ${\Delta\nu}$ has been measured with high confidence and found to be incommensurable with $\nu_{\rm osc}$. Commensurability of $\nu_{\rm osc}$ and ${\Delta\nu}$ appears to be a general feature of these oscillations.

{\em Low-frequency QPOs}.---Three types of QPOs or power spectral bumps with frequencies $\la\,$100~Hz have been detected in many neutron star sources (see van der Klis 1989, 2000): (1)~the $\sim\,$6--20~Hz normal- and flaring-branch oscillations observed in power spectra of the X-ray emission from the so-called Z sources when they are in their normal and flaring branch spectral states, (2)~the $\sim\,$15--60~Hz horizontal branch QPO seen in the Z sources when they are in their horizontal-branch spectral state, and (3)~the $\sim\,$20--80~Hz peaked noise components and low-frequency QPOs (LFQPOs) observed in the atoll sources. The latter two types of QPOs may be related to one another. The frequencies of these QPOs and other power spectral features are positively correlated with the X-ray flux and the frequencies of the kilohertz QPOs (van Straaten et al.\ 2001).

\subsection{Brightness Oscillations in Black Hole Candidates}
\label{sec:BH-Osc}

A variety of QPOs have been detected in black hole candidates (see Remillard et al.\ 2001; van der Klis 1998, 2000). QPOs with frequencies $\ga\,$50~Hz have been detected in half a dozen black hole candidates. In addition, power spectra of the X-ray brightness of several black hole candidates and the peculiar source \hbox{Cir~X-1} show two QPOs or broad noise peaks with frequencies between $\sim\,$0.1~Hz and 10~Hz.

\begin{table}[t]
\begin{center}
\begin{minipage}{122mm}
\caption{$\!\!\!\!$High-Frequency QPO Pairs in Black Hole Candidates}
\label{table.BH-HFQPOs}
\vspace{3pt}
\begin{tabular}{lccccc}
\tableline\tableline\noalign{\kern 4pt}
Source   &$\nu$~(Hz)  &rms amplitude  &$\nu/{\rm FWHM}$  &${\Delta\nu}/\nu$$^a$  &Ref.\\
\noalign{\kern 3pt}
\tableline
\noalign{\kern 3pt}
\hbox{GRS~1915$+$105}   &40    &$\la 2$\%   &$\sim\,5$--8   &$\la 1$\%    &1\\  
\noalign{\kern 1pt}
&67    &$\sim 1$--2\%  &$\sim 6$--20  &$\sim\,5$\%  &2,3\\
\noalign{\kern 1pt}
\hbox{GRO~J1655$-$40}   &300    &$\sim 0.8$\%  &$\sim\,4$   &$\la 10$\%  &4\\
\noalign{\kern 1pt}
&450   &$\sim 5$\%  &\ \ \ 23   &$\la 4$\%   &5\\
\noalign{\kern 4pt}
\tableline
\noalign{\kern 4pt}
\end{tabular}
$^a$${\Delta\nu}$ is the observed variation in the QPO centroid frequency $\nu$ or the upper limit on this variation. References: [1]~Strohmayer (2001d)  [2]~Morgan et al.\ 1997  [3]~Remillard \& Morgan 1998  [4]~Remillard et al.\ 1999  [5]~Strohmayer (2001c) 
\end{minipage}
\end{center}
\end{table}

{\em High-frequency QPOs}.---Pairs of high-frequency QPOs with relatively stable frequencies have been detected in two black-hole systems (see Table~\ref{table.BH-HFQPOs}). QPOs at 40~Hz (Strohmayer 2001d) and 67~Hz (Morgan, Remillard, \& Greiner 1997; Remillard \& Morgan 1998) have been seen in \hbox{GRS~1915$+$105}. Their frequencies remain almost constant as the X-ray flux changes (any change in the frequency of the 40~Hz oscillation during the sequence of observations must have been $\la 1$\%; the frequency of the 67~Hz oscillation varies by only a few percent as the X-ray flux varies by a factor of several). Stable QPOs at 300~Hz (Remillard et al.\ 1999) and 450~Hz (Strohmayer 2001c) have been observed in \hbox{GRO~J1655$-$40}.   

Single, relatively high-frequency QPOs or peaked noise features have been detected in \hbox{XTE~J1550$-$564} (Remillard et al.\ 1999a; Homan et al.\ 2001; Miller et al.\ 2001), \hbox{4U~1630$-$47} (Remillard \& Morgan 1999), \hbox{XTE~J1859$+$226} (Cui et al.\ 2000), and \hbox{Cyg~X-1} (Remillard et al.\ 1999a, 1999b). The frequencies of some of these high-frequency QPOs are known to vary substantially; others have not been seen to vary but have been observed only briefly.

{\em Low-frequency QPOs}.---A variety low-frequency QPOs or peaked noise components have been identified in the power spectra of black hole candidates (see Psaltis, Belloni, \& van der Klis 1999; van der Klis 2000).

\subsection {High-Frequency QPO Pairs in Neutron Stars and Black Holes}

The discovery of pairs of high-frequency QPOs in the two black hole sources \hbox{GRO~J1655$-$40} and \hbox{GRS~1915$+$105} raises the question whether they are produced by the same physical mechanism as the kilohertz QPO pairs seen in more than twenty neutron star sources (Strohmayer 2001d). The frequencies of these QPOs are roughly in the range expected for dynamical frequencies in the vicinity of stellar-mass black holes and $\sim 2M_{\sun}$ neutron stars, respectively.

One indication that the high-frequency QPO pairs seen in the black holes have a different origin from the kilohertz QPO pairs observed in neutron stars is that their phenomenology is significantly different (see van der Klis 2000; Strohmayer 2001c, 2001d). Perhaps most significantly, the frequencies of the kilohertz QPOs increase by factors $\sim\,$2--3 in some sources when the X-ray flux increases. In contrast, the frequencies of the high-frequency black hole QPOs do not change much if at all when the X-ray flux increases. Pairs of kilohertz QPOs are seen in both the Z and atoll sources, even though the latter sources are up to $\sim 300$ times less luminous than the former. In contrast, the 300~Hz QPO in \hbox{GRO~J1655$-$40} is detected when the source is flaring. 

The rms relative amplitudes of the kilohertz QPOs in the 2--60~keV energy band are sometimes much greater than their FWHM/$\nu_{\rm QPO}$ ratios. This excludes models in which the QPO is produced by oscillations in the luminosity of gas in a ring in the disk that is narrow enough so that the range of orbital frequencies in the ring does not exceed the FWHM of the QPO. In contrast, the FWHM/$\nu_{\rm QPO}$ ratios of the high-frequency black hole QPOs are generally much larger than those of the kilohertz QPOs and all exceed the rms relative amplitude of the respective QPOs. Finally, the rms relative amplitudes of both kilohertz QPOs increase with X-ray energy up to at least $\sim 12$--15~keV. The rms amplitude of the 300~Hz QPO in \hbox{GRO~J1655$-$40} is stronger in the soft X-ray band whereas the 450~Hz QPO is detected only above $\sim 13$~keV; conversely, the 67~Hz QPO in \hbox{GRS~1915$+$105} is detected in the soft X-ray band, whereas the 40~Hz QPO is only detected at energies above $\sim 13$~keV. The question of whether the high-frequency QPO pairs seen in neutron star and black holes sources are physically related is discussed further in \S3.4. This issue clearly deserves further attention.

\section{QPO Models and Comparisons with Observation}
\label{sec:mechanisms}

A variety of models have been proposed to explain the various QPOs observed in neutron star and black hole LMXBs. In this section I describe several QPO mechanisms that have been proposed and compare them with some of the data on QPOs and other power spectral features in neutron star and black hole sources. I then discuss briefly our general understanding of QPOs in neutron stars and black holes and their relationships to one another.

\subsection{QPO Models}

{\em Sonic-point beat-frequency (SPBF) model}.--- General relativistic gas dynamical and radiation transport calculations (Miller \& Lamb 1993, 1996; see also Tsuribe et al.\ 1995 and references therein) show that at low accretion rates, the azimuthal drag force exerted by radiation from a relativistic star on gas orbiting near it removes angular momentum from the gas sufficiently quickly that the gas accelerates radially inward, causing the radial flow to become supersonic before it reaches the star or the ISCO, if one exists. This is an effect of strong-field general relativity (a transition of this type is not possible in Newtonian gravity). To lowest order, the radius $r_{\rm sp}$ of the sonic point is determined by balancing the angular momentum brought into the interaction region by the accretion flow, which is $\propto {\dot M} \ell_{\rm acc}$, where $\ell_{\rm acc}$ is the mean specific angular momentum of the accreting gas, and the angular momentum removed from the flow by the radiation drag, which is $\propto L_{\rm acc} \propto {\dot M}$. Hence to lowest order, $r_{\rm sp}$ is independent of ${\dot M}$. However, the optical depth of the flow from the disk to the stellar surface increases with ${\dot M}$, reducing the radiation drag as ${\dot M}$ increases. This causes $r_{\rm sp}$ to decrease with increasing ${\dot M}$. At sufficiently high ${\dot M}$, the accretion flow remains nearly circular until it reaches the ISCO, if one exists. Once it reaches the ISCO, gas accelerates inward and becomes supersonic even if it does not lose any angular momentum to the radiation field and hence $r_{\rm sp} \approx r_{\rm isco} = {\rm const}$. The existence of a sonic point is an effect of the strong gravitational field when $r_{\rm sp} \approx r_{\rm isco}$ as well as when $r_{\rm sp} > r_{\rm isco}$. If the neutron star is so large that there is no ISCO, the region of nearly circular flow may extend to the stellar surface.

Miller, Lamb, \& Psaltis (1998b, 1998c) and Lamb, Miller, \& Psaltis (1998b) showed that inhomogeneities in the accretion flow near $r_{\rm sp}$ produce two related high-frequency QPOs. Where gas falling inward from clumps near $r_{\rm sp}$ impacts the stellar surface, it produces bright footprints that move around the star with a frequency nearly equal to the general relativistic orbital frequency $\nu_{\rm orb}$ of the clumps at $r_{\rm sp}$, causing the X-ray flux seen by distant observers to oscillate as the footprint moves into and out of view. The frequency of this oscillation is usually less than the frequency $\nu_{Kc}(r_{\rm sp})$ of a general relativistic Keplerian circular orbit at $r_{\rm sp}$, because the radial component of the radiation force causes $\nu_{\rm orb}$ to be less than $\nu_{Kc}(r_{\rm sp})$ and the radially inward drift of the clumps causes the footprint to move around the star with a frequency less than $\nu_{\rm orb}$ (Lamb \& Miller 2001).

The weak magnetic field of the star funnels extra gas toward the magnetic poles, producing a broad beam of slightly stronger radiation that rotates with the star. For disk flows that are prograde relative to the star's spin, clumps orbiting near $r_{\rm sp}$ move through this beam with frequency $\nu_B \equiv \nu_{\rm orb} - \nu_s$. When a clump is illuminated by the beam, gas in it loses angular momentum to the radiation at a faster rate, so more gas falls inward, causing the luminosity of the footprint to oscillate with a frequency close to $\nu_B$. The inward radial velocity of the flow usually makes the frequency slightly greater than $\nu_B$ (Lamb \& Miller 2001). If the disk flow near the star is temporarily retrograde, clumps orbiting near $r_{\rm sp}$ move through the beam with frequency $\nu_B' \equiv \nu_{\rm orb} + \nu_s$, causing the luminosity to oscillate with a frequency close to but usually greater than $\nu_B'$.

Light travel time effects are negligible and hence the time between two successive maxima of the mass inflow rate from a given clump is very close to the beat period ${\Delta t}_{\rm B} \equiv 1/(\nu_{\rm orb} - \nu_s)$. Let $t_1$ and $t_2 = t_1 + {\Delta t}_{\rm B}$ be the times of successive maxima in the mass inflow rate. Then the frequencies of the lower and upper kilohertz QPOs are
$$
\nu_1={1\over{t_2-t_1+(\Delta t_2-\Delta t_1)}}
\qquad  {\rm and} \qquad
\nu_2={\phi_2-\phi_1+(\Delta\phi_2-\Delta\phi_1)\over{
t_2-t_1+(\Delta t_2-\Delta t_1)}} \;,
$$
where $2\pi\phi_1$ and $2\pi\phi_2$ are the azimuthal coordinates of the clump at $t_1$ and $t_2$, ${\Delta t}_1$ and ${\Delta t}_2$ are the times required for the gas stripped from the clump at $t_1$ and $t_2$ to reach the surface of the star, and $2\pi{\Delta\phi}_1$ and $2\pi{\Delta\phi}_2$ are the changes in the azimuthal phase of the gas released at $t_1$ and $t_2$ as it falls from the clump to the stellar surface.

The inward drift of a given clump causes the time required for gas to spiral inward from it to the stellar surface to decrease steadily. Hence ${\Delta t}_2 < {\Delta t}_1$. Disk flow is usually expected to be prograde relative to the star's spin. Then, to first order in the inward drift velocity $v_{\rm cl}$ of a clump, $\nu_1$ is $\nu_{\rm B} (1+v_{\rm cl} \partial_r\Delta t) > \nu_{\rm B}$; $\nu_1$ is greater than $\nu_{\rm B}$ because the inward motion of the clump reduces the spatial separation between successive maxima of the mass flow rate, analogous to the Doppler shift of a sound wave. Numerical simulations (Lamb \& Miller 2001) show that $\nu_2$ is usually less than $\nu_{\rm orb}$, because the gas falling inward from the clump winds a progressively smaller distance around the star as the clump drifts inward. Hence ${\Delta\nu}$ is less than $\nu_s$. If instead the accretion flow near the star is temporarily retrograde, $\nu_1$ is expected to be comparable to but less than $\nu_{\rm orb}$, while $\nu_2$ is usually comparable to but greater than $\nu_{\rm Bp} = \nu_{\rm orb} + \nu_s$. Hence in this case ${\Delta\nu}$ is usually {\em greater\/} than $\nu_s$.

In the SPBF model, the frequencies of the X-ray oscillations depend on the mass $M$ and angular momentum $J$ of the star, the EOS of neutron star matter, the magnetic field of the star, and ${\dot M}$. Numerical simulations (Miller et al.\ 1998b; Lamb \& Miller 2001) indicate that in its simplest form, the SPBF model gives $r_{\rm sp} \sim 1$--$3\,R_{\rm star}$ and X-ray oscillation frequencies $\sim 500$--1500~Hz. Deviations of $\nu_1$ and $\nu_2$ from $\nu_B$ and $\nu_{\rm orb}$ by $\la 5$\% caused by the inward motion of the accretion flow at $r_{\rm sp}$ are sufficient to account for the observed differences between ${\Delta\nu}$ and $\nu_s$ and the decrease of ${\Delta\nu}$ with increasing $\nu_2$ described in \S\ref{sec:NS-Osc} The X-ray flux for both kilohertz QPOs is produced chiefly at the stellar surface, where most of the gravitational energy is released, and hence both can have relatively high rms amplitudes. The existence of a sonic point in the flow is a strong-field general relativistic effect, and the frequencies of the sonic-point kilohertz QPOs are governed primarily by the spacetime near the neutron star. Hence they are sensitive probes of this spacetime, including whether there is an innermost stable circular orbit (ISCO) and the gravitomagnetic torque created by the spin of the star. In the SPBF model, the low-frequency QPOs are assumed to be produced by the magnetospheric beat-frequency mechanism described below.

{\em Relativistic precession (RP) models}.---These models identify the frequencies of QPOs with the basic frequencies of free-particle geodesic motion around the compact object. The frequencies considered are the general relativistic mean azimuthal (Keplerian) frequency $\nu_K$, radial epicyclic frequency $\nu_r$, and meridional oscillation frequency $\nu_\psi$. For circular geodesics around a Kerr black hole of mass $M$, these frequencies (in units of $c^3\!/GM$) are ${\hat\nu}_K(M,j,r) = [2\pi{\hat r}^{3/2} (1+j{\hat r}^{-3/2})]^{-1}$, ${\hat\nu}_r(M,j,r) = |{\hat\nu}_K|(1-6{\hat r}^{-1} + 8j{\hat r}^{-3/2} -3j^2{\hat r}^{-2})^{1/2}$, and ${\hat\nu}_\psi(M,j,r) = |{\hat\nu}_K|(1-4j{\hat r}^{-3/2} + 3j^2 {\hat r}^{-2})^{1/2}$, where ${\hat r}$ is the Boyer-Lindquist radius in units of $GM/c^2$ and $j \equiv cJ/GM^2$ (see Markovi\'c 2000). The frequencies of eccentric geodesics around black holes cannot be expressed analytically. For neutron stars, even the frequencies of circular geodesics must be computed numerically, using general relativistic stellar models.

In the clump precession model (Stella \& Vietri 1998, 1999; Cui, Zhang, \& Chen 1998b; Karas 1999; Stella, Vietri, \& Morsink 1999), the frequencies of QPOs and other features in the  power spectra of neutron stars and black holes are the frequencies of purely geodesic motion of gas clumps around the compact object. The upper and lower kilohertz QPO frequencies are assumed to be the Keplerian frequency $\nu_K$ and the apsidal precession frequency $\nu_{\rm AP} = \nu_K - \nu_r$; the frequency of any low-frequency QPO is assumed to be the nodal precession frequency $\nu_{\rm NP} = \nu_K - \nu_\psi$ or $2\,\nu_{\rm NP}$. These frequencies depend on $M$, $j$, and the periastron and apastron radii $r_p$ and $r_a$ of the geodesic; for neutron stars, they also depend on the EOS of neutron star matter.

In the ring precession model (Psaltis \& Norman 2000), gas in a narrow ring at a special radius in the disk is assumed to behave as a separate entity. A ring with the properties proposed has a dense spectrum of resonances, including resonances near the frequencies $2\,\nu_{\rm NP}$, $\nu_{\rm AP}$, and $\nu_K$ of an infinitesimally eccentric geodesic at the radius of the ring (motion along any geodesic with a finite eccentricity is quickly damped; see Markovi\'c \& Lamb 2000). In the ring precession model, these three frequencies are assumed to be the dominant X-ray oscillation frequencies and the frequencies of the low-frequency and kilohertz QPOs. These frequencies depend on $M$, $j$, and the radius $r$ of the nearly circular geodesic; for neutron stars, they also depend on the EOS of neutron star matter.

At present, the physics that determines which geodesics are populated by clumps in the precessing clump model or the radius of the ring in the precessing ring model has not been specified. Hence, in their present form these models make no definite predictions of the frequencies that should be seen or whether they depend on the accretion rate and luminosity.

{\em Disk oscillation models}.---These ``diskoseismic'' models associate QPOs with the oscillation modes of the inner accretion disk (see Kato \& Fukue 1980; Kato 1990, 1993; Ipser \& Lindblom 1991, 1992; Kato \& Honma 1991; Ipser 1994, 1996; for reviews, see Kato, Fukue, \& Mineshige 1998; Markovi\'c \& Lamb 1998; Wagoner 1998). Much of this work has focused on the low-order modes of geometrically thin, pseudo-Newtonian or relativistic (Novikov \& Thorne 1973; Page \& Thorne 1974) disks around black holes, but there has been some work on disks around neutron stars. 
The frequencies of the low-order $g$, $c$, and $p$-modes depend on the mass and angular momentum of the compact object but hardly at all on the accretion rate, because they are trapped waves that exist only near the ISCO. In studies of these modes, viscous energy dissipation is considered in computing the vertical structure of the disk, but angular momentum transfer by the viscous shear stress has usually been neglected. In this approximation, the motion of the fluid in the unperturbed disk is circular, i.e., there is no accretion. The $w$-modes of the inner disk have been studied by treating the disk as a succession of rings that exchange angular momentum via advection and the viscous shear stress. The frequencies of the $w$-modes depend on where the nearly circular Keplerian flow ends and can vary substantially. 

The gravity-inertial modes of geometrically thin, relativistic disks are standing waves trapped in the region of the disk where their frequencies are $\le\nu_r$ (see Wagoner, Silbergleit, \& Ortega-Rodr\'{\i}guez 2001). This trapping is a general relativistic effect. Hence $g$-modes exist only close to the compact object. The axisymmetric ($m = 0$) $g$-modes are centered at the radius ${\hat r}_g$ where $\nu_r$ reaches its maximum value and extend over a radial interval ${\Delta\hat r}_g \approx 3.3\, (L/L_E)^{1/2}$, where $L_E$ is the Eddington critical luminosity; ${\hat r}_g$ depends on $j$, decreasing from 8 for $j=0$ to $\approx3$ for $j=0.95$. The restoring force that causes oscillatory motion is primarily the net gravitational-centrifugal force (see Perez et al.\ 1997). The frequencies $\nu_g$ of the $g$-modes that have only a few nodes in the vertical and radial directions are close to $\nu_r(r_g)$, depend only weakly on the radial ($n$) and vertical ($l$) mode numbers, and decrease by $\la5$\% when the luminosity doubles (Wagoner et al.\ 2001).

The fundamental ($|m| = l = 1$) corrugation modes of geometrically thin, relativistic disks are trapped waves that displace the disk in the vertical direction and rotate around the compact object (see Silbergleit, Wagoner, \& Ortega-Rodr\'{\i}guez 2001; Wagoner et al.\ 2001). Trapping of these waves is also a general relativistic effect. The fundamental $c$-modes are trapped between $r_{\rm isco}$ and the outer trapping radius $r_c$. For $n=0$, $M \sim 10M_{\sun}$, and $L \sim 0.1\,L_E$, $r_c \approx r_{\rm isco} + 0.09\, j^{-0.8} (1-j)^{0.2}$. Hence, only if the compact object is slowly rotating do these modes extend over a sizeable radial interval. In the low-$j$ limit, the frequency of the fundamental $c$-modes is equal to the gravitomagnetic frequency ${\hat\nu}_{\rm gm}(r) = {\hat\nu}_K(r) - {\hat\nu}_\psi(r) = 4\pi j\nu_K(r)^2$ at the outer boundary of the trapping region, i.e., $ {\hat\nu}_c = {\hat\nu}_{\rm gm}(r_c)$; $ {\hat\nu}_c$ decreases slightly with increasing luminosity.

Geometrically thin, viscous accretion disks around spinning compact objects have a family of high-frequency warping modes that are localized near the inner edge of the near-Keplerian flow (Markovi\'c \& Lamb 1998; see also Armitage \& Natarajan 1999). These $w$-modes are spiral corrugations that rotate around the compact object with a frequency close to the gravitomagnetic (frame-dragging) frequency ${\hat\nu}_{\rm gm}$ at the radius where their amplitudes are largest. The shapes of these $w$-modes are determined by competition between the gravitomagnetic and viscous torques acting on the gas in the disk. For black holes, the high-frequency $w$-modes are situated just outside the ISCO. For neutron stars, they may be situated just outside the stellar surface, ISCO, sonic radius, or magnetospheric boundary, depending on which effect terminates the nearly circular Keplerian flow. The highest-frequency $w$-modes in this family are closely spaced in frequency and weakly damped ($Q$-values up to $\sim\,50$). They are distinct from but related to the low-frequency $w$-modes, which only perturb the accretion disk far from the central compact object (see Maloney, Begelman, \& Pringle 1996; Maloney, Begelman, \& Nowak 1998; Pringle 1996; Markovi\'c \& Lamb 1998).

Accretion disks also have high-frequency $p$-modes, which are restored primarily by pressure-gradient forces. The fundamental ($|m| = l = 0$) $p$-modes span only a very small radial interval just outside $r_{isco}$ and have frequencies $|{\hat\nu}_p| \ge {\hat\nu}_r$ (Wagoner et al.\ 2001). Their frequencies increase by $\sim\,$5\% when the luminosity doubles.

There are a variety of mechanisms by which these disk modes could modulate the X-ray flux of an accreting neutron star or black hole (see, e.g., Nowak \& Wagoner 1993; Nowak et al.\ 1997).

{\em Magnetospheric beat-frequency (MBF) model}.---This model generates a relatively coherent QPO by the interaction of the stellar magnetic field with gas in the accretion disk (Alpar \& Shaham 1985; Lamb et al.\ 1985;  Lamb 1991; Miller et al.\ 1998b). The frequency of the QPO is the rate at which the orientation of the stellar magnetic field repeats, relative to the inhomogeneities in the disk at the coupling radius. This frequency is the magnetospheric beat frequency $\nu_{\rm Bm,k} = k(\nu_{\rm orb,m} - \nu_s)$, where $\nu_{\rm orb,m}$ is the orbital frequency at the radius $r_m$ at which the gas in the disk couples strongly to the stellar magnetic field and $k =  1,2,\ldots$ describes the symmetry of the magnetic field there (e.g., for an offset dipole, both the $k=1$ and $k=2$ beat frequencies would be generated). The frequency of the QPO depends on $M$, $j$, the neutron star matter equation of state (EOS), ${\dot M}$, the disk structure, and the important multipole moments of the stellar magnetic field at $r_m$. 

{\em Coronal flow oscillation model}.---Spectral modeling suggests that a central region of radially inflowing gas forms around neutron stars in LMXBs when their accretion rates approach the Eddington critical rate ${\dot M}_E$ (Lamb 1989, 1991). Numerical studies of such flows show that they become overstable when the accretion rate reaches a critical value somewhat less than ${\dot M}_E$ (Fortner, Lamb, \& Miller 1989; Fortner 1992). The resulting oscillation modulates the X-ray spectrum at the frequency $\nu_{\rm osc,c} \approx 2/t_f \approx 6$~Hz, where $t_f$ is the inflow time from the outer edge of the radially inflowing gas. This edge is determined by the interaction of the radiation from the neutron star with the disk corona.

\subsection{Comparison of Models with Neutron Star QPOs}

As noted in \S2.1, a key feature of all the neutron star QPOs is that their frequencies vary significantly and are positively correlated with each other and with the inferred X-ray flux. This imposes very significant constraints on models of the neutron star QPOs. The frequency $\nu_2$ of the upper kilohertz QPO is comparable to the frequencies of orbits near the ISCO or the surface of a 1.5--$2M_{\sun}$ neutron star. Most models assume that $\nu_2$ is the orbital frequency $\nu_{\rm orb}$ of gas at some special radius in the accretion disk (see Lamb et al.\ 1998b; van der Klis 1998, 2000). A fundamental question in such models is, what physics determines this special radius? Whatever it is, it must cause the radius to vary by a factor $\sim\,$1.5 as the luminosity of the source varies by $\sim\,$50\%.

{\em Sonic-point beat-frequency model}.---In the SPBF model, the special radius in the disk is the radius $r_{\rm sp}$ of the sonic transition produced either by interaction of the radiation from the neutron star with the accretion flow or by the approach of the flow to the ISCO (see \S3.1). The model predicts that the upper and lower kilohertz QPOs usually have frequencies close to $\nu_{\rm orb}(r_{\rm sp})$ and $\nu_{\rm orb}(r_{\rm sp}) - \nu_s$, respectively. The strongest motivation for the SPBF model is the evidence that the burst oscillations have frequencies close to the stellar spin frequency or its second harmonic (see \S2.1) and that the frequency separation ${\Delta\nu} \equiv \nu_2 - \nu_1$ is close to $\nu_{\rm osc}$ or $0.5\nu_{\rm osc}$ (see Table~\ref{table.BurstOsc}). The steady accumulation of sources in which ${\Delta\nu}$ and $\nu_{\rm osc}$ are commensurable and the absence of any that are not makes it seem unlikely that this commensurability is merely a coincidence.

As the accretion rate ${\dot M}_d$ through the inner disk increases, $r_{\rm sp}$ decreases, causing both kilohertz QPO frequencies to increase, in agreement with their observed behavior. In addition to explaining the observed frequency ranges of $\nu_1$ and $\nu_2$ and their increase with inferred ${\dot M}_d$, the SPBF model appears able to account for the relatively high amplitudes and coherence of the kilohertz QPOs and the steep increase of QPO amplitude with photon energy (Miller et al.\ 1998b). The amplitudes of the kilohertz QPOs and the amplitude and frequency behavior of the HBOs suggest that the latter are produced by the magnetospheric beat-frequency mechanism (see Miller et al.\ 1998b; Psaltis et al.\ 1999b). The neutron star magnetic fields indicated are comparable to those expected if the kilohertz QPO sources are the progenitors of the recycled millisecond pulsars (Miller et al.\ 1998b; Lorimer 2000).

 \begin{figure}[t]   
 \vglue -2.2 truecm 
 \centerline{\epsfxsize=11.9 truecm \hskip +0.1 truecm
 \epsfbox{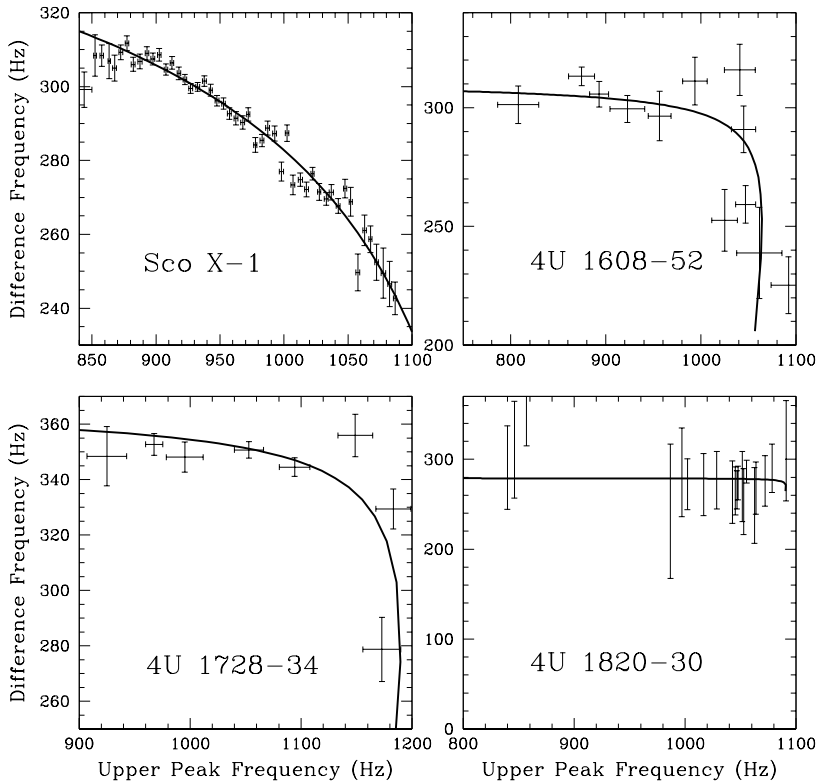}}
 \vglue -3.3 truecm
 \caption{Fits of the sonic-point beat-frequency gas dynamical model to the kilohertz QPO frequencies observed in four sources, assuming the disk flow is prograde relative to the neutron star; the spacetime was approximated by the Schwarzschild spacetime. The best-fitting neutron star masses and spin rates and the resulting $\chi^2$ values are listed in Table~\ref{table.NSProperties} below. After of Lamb \& Miller (2001).}
\label{fig:SPBFComparison}
\end{figure}

Stella \& Vietri (1998) suggested that the kilohertz QPOs are generated by the SPBF mechanism but that the HBOs and low-frequency QPOs are produced by nodal precession of gas at $r_{\rm sp}$. However, further study has shown that the frequencies and frequency behaviors of the HBOs and LFQPOs are inconsistent with nodal precession, unless the spin rates of these stars are $\sim\,$800--900~Hz, much larger than the rates indicated by the burst oscillations and inconsistent with the SPBF model (Markovi\'c \& Lamb 1998; Psaltis et al.\ 1999; Morsink \& Stella 1999; Kalogera \& Psaltis 1999). 

As explained in \S2.1, the SPBF model predicts that ${\Delta\nu}$ is usually somewhat less than $\nu_{s}$ and decreases with increasing ${\dot M}_d$ (in Miller et al.\ 1998b the inward radial velocity of the disk flow was neglected, in which case ${\Delta\nu} = \nu_{s}$). The inflow velocities expected at $r_{\rm sp}$ naturally produce shifts in $\nu_1$ and $\nu_2$ of the size (0.2\%--5\%) needed to explain the differences observed between ${\Delta\nu}$ and $\nu_{\rm osc}$ and the variation of ${\Delta\nu}$ with $\nu_2$. Figure~\ref{fig:SPBFComparison} shows that fits of the SPBF model agree fairly well with the observed behavior of the kilohertz QPOs in the four sources shown.

\begin{figure}   
\vglue-0 truecm
\hbox{\hglue+0.8 truecm
{\epsfysize=2.0 truein\epsfbox{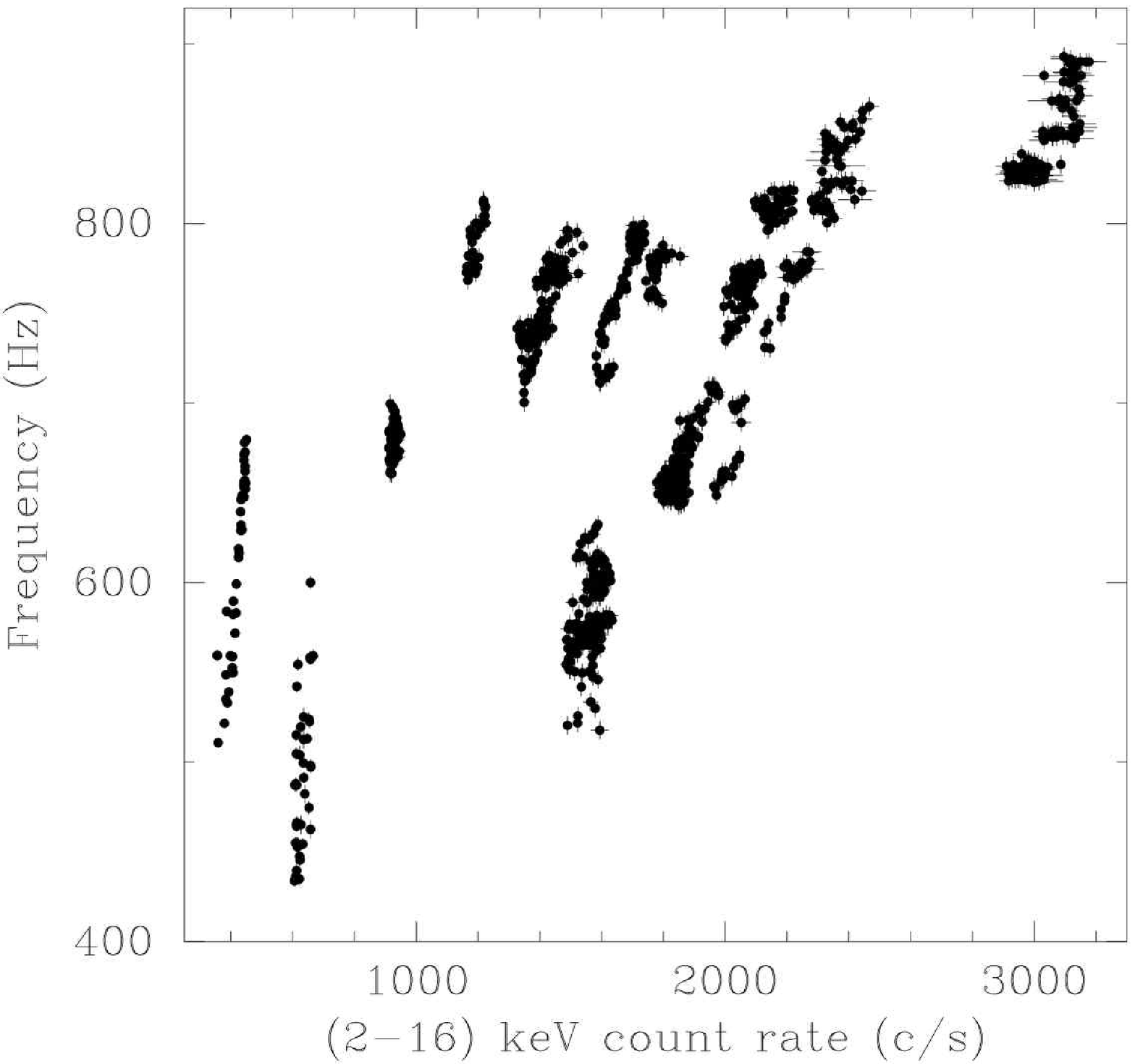}}}
\vglue-5.15 truecm
\hbox{\hglue+6.8 truecm
{\epsfysize=2.08 truein\epsfbox{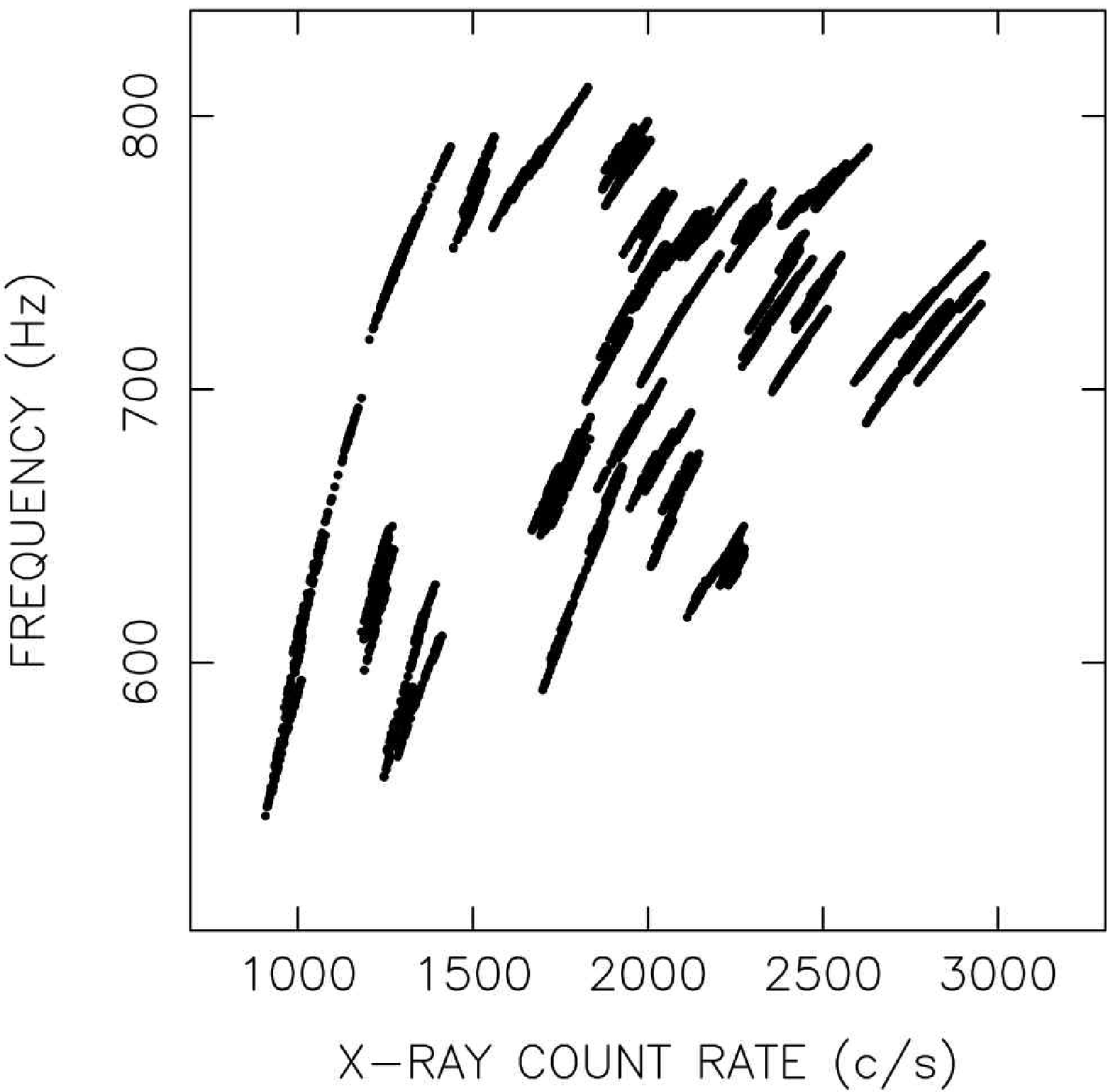}}}
\vglue+0 truecm
\caption{{\em Left\/}: $\nu_1$ vs.\ countrate tracks observed in \hbox{4U~11608$-$52} (M\'endez et al.\ 1999). {\em Right\/}: Simulated tracks (van der Klis 2001).}
 \label{fig:1608-tracks}
 \end{figure}

One puzzling aspect of the kilohertz QPOs is that $\nu_1$ and $\nu_2$ vary differently with countrate (or X-ray flux) on different timescales. On short timescales (typically $\sim\,$hours), the frequencies of the kilohertz QPOs are tightly correlated with the apparent X-ray luminosity, producing relatively narrow, sloping tracks in plots of $\nu_{\rm QPO}$ vs.\ flux, but on longer timescales (typically $\sim\,$days), these tracks drift in flux by as much as 50\% (see the left panel of Fig.~\ref{fig:1608-tracks}). Understanding the cause of these drifts is important in its own right, but it is essential in order to determine whether the plateaus seen in these tracks are evidence for the existence of an ISCO (see \S4).

In the sonic-point model, $r_{\rm sp}$ is determined by a balance between the inward flux of angular momentum through the inner disk, which is proportional to ${\dot M}_d$, and the radiation drag on the gas there, which is proportional to the luminosity. Using a phenomenological model, van der Klis (2001) has shown that this mechanism may be able to explain the drift of the $\nu_1$ and $\nu_2$ vs.\ countrate tracks, if the luminosity---and hence the radiation drag---has a component that is proportional to the {\em average\/} of the {\em future\/} mass flux through the inner disk, with an averaging timescale of about a day, as might occur if part of the mass flux onto the star comes from rapid infall of gas from the disk at radii $> r_{\rm sp}$ (see the right panel of Fig.~\ref{fig:1608-tracks}). This possibility clearly deserves further study.

Probably the most significant issue for the SPBF model is the lack of compelling evidence for dynamically significant magnetic fields in these neutron stars. There is a wealth of indirect evidence for such fields: disk accretion would spin stars up to spin rates much greater than $\sim\,350$~Hz if their magnetic fields were negligible (Ghosh \& Lamb 1992); it is difficult to understand the high stability of the oscillations in burst tails if there is no marker on the surface of the star; the requirement that many neutron stars in LMXBs must have dipole magnetic fields $\sim 10^8$--$10^9$~G, if they are the progenitors of the recycled millisecond pulsars (Kulkarni \&Phinney 1994; Lorimer 2001); the correlation of X-ray spectra and HBO, LFQPO, and kilohertz QPO properties with inferred magnetic field strength (see Psaltis, Lamb, \& Miller 1995, 1998a; Miller et al.\ 1998b; Lamb et al.\ 1998b, Fig.~2); and other evidence. However, this evidence is not considered compelling by some.

The discovery of burst oscillations in \hbox{SAX~J1808$-$369} with a frequency close to its 400~Hz frequency spin frequency (see \S2.1) is consistent with the hypothesis that X-ray burst oscillations are generated by the spin of the star. However, no kilohertz QPOs were detected during the brief interval in 1998 when \hbox{SAX~J1808$-$369} was observed by {\em RXTE}, so its relation to the kilohertz QPO sources remains unclear at present. Continuous observations of this source should be a high priority when it next appears. Another issue for the SPBF model is the apparent similarity of the correlations between $\nu_{\rm LFQPO}$ and $\nu_1$ in neutron stars and between the frequencies of some of the low-frequency band-limited noise components seen in the power spectra of black holes (see \S3.4).

{\em Magnetospheric and magnetospheric beat-frequency models}.---The possibility that the frequencies of the kilohertz QPOs might be the Keplerian frequency $\nu_{\rm Km}$ at the radius $r_m$ where the gas in the disk strongly couples to the star's magnetic field and the magnetospheric beat frequency $\nu_{Bm} \equiv \nu_{\rm Km} - \nu_s$ has been suggested by Strohmayer et al.\ (1996) and Ford et al.\ (1997). Cui et al.\ (1998), Cui (2000), and Campana (2000) have also discussed the possibility that QPOs are generated at $r_m$. However, it is difficult to understand the production of two fairly coherent oscillations, the large observed variation of the kilohertz QPO frequencies, and other properties of the kilohertz QPOs in this model (see Miller et al.\ 1998b; Lamb et al.\ 1998a, 1998b).

{\em Relativistic precession models}.---In the clump precession model, the kilohertz QPO frequencies $\nu_2$ and $\nu_1$ are assumed to be the azimuthal and apsidal precession frequencies $\nu_{\rm K}$ and $\nu_{\rm AP}$ of gas clumps on slightly eccentric, tilted geodesics, while the frequencies of the LFQPO and HBO are the first or second harmonics of the nodal precession frequencies $\nu_{\rm NP}$ of these geodesics. As noted in \S3.1, the physics that determines how the clumps are formed and which geodesics are populated has not been specified. Hence the model makes no definite predictions about whether the frequencies of the QPOs depend on the accretion rate.

\begin{figure}[t]   
\vglue-0 truecm
\hbox{\hglue+.65 truecm
{\epsfysize=2.75truein\epsfbox{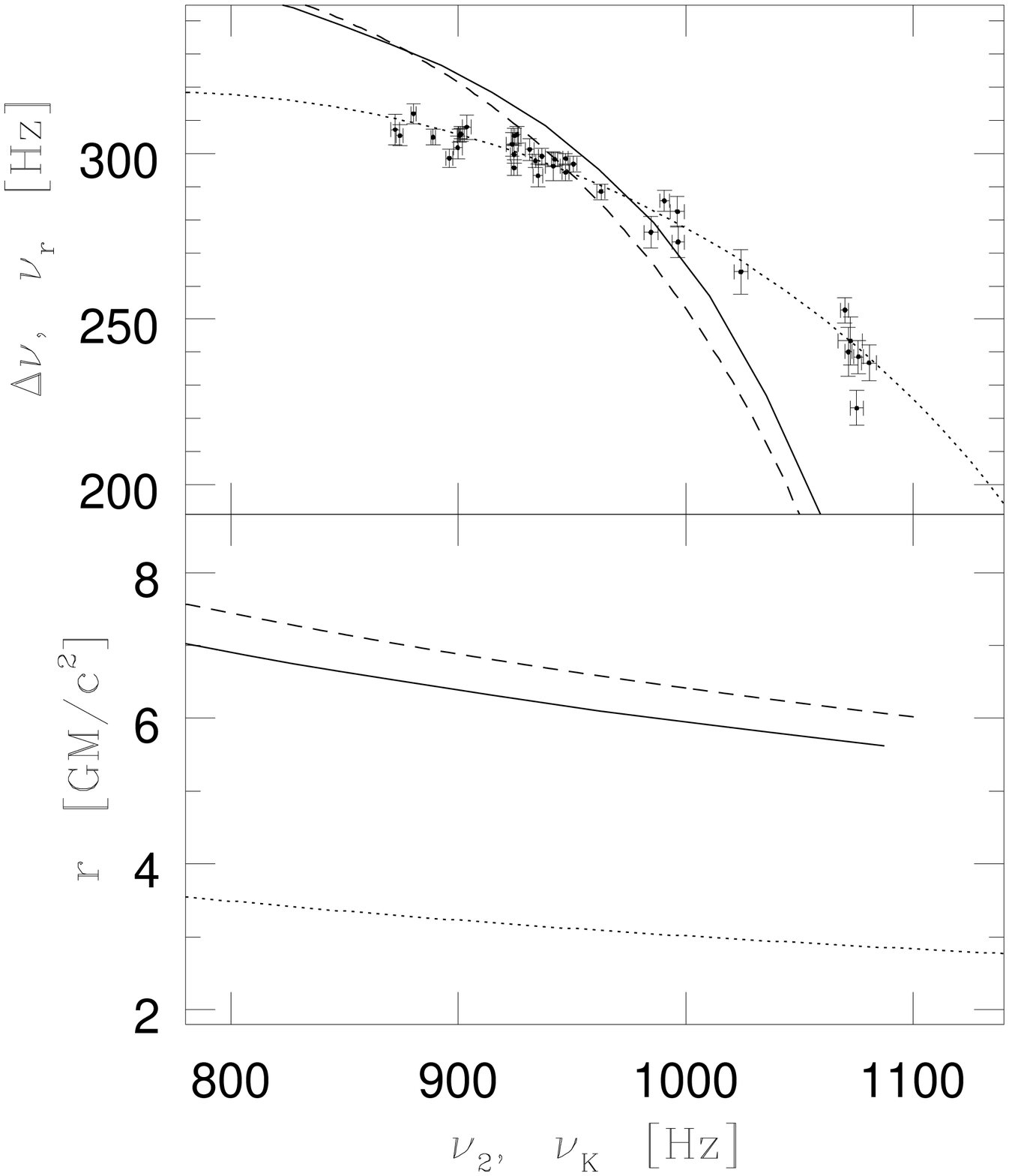}}
\hglue 0.25 truecm
{\epsfysize=2.75truein\epsfbox{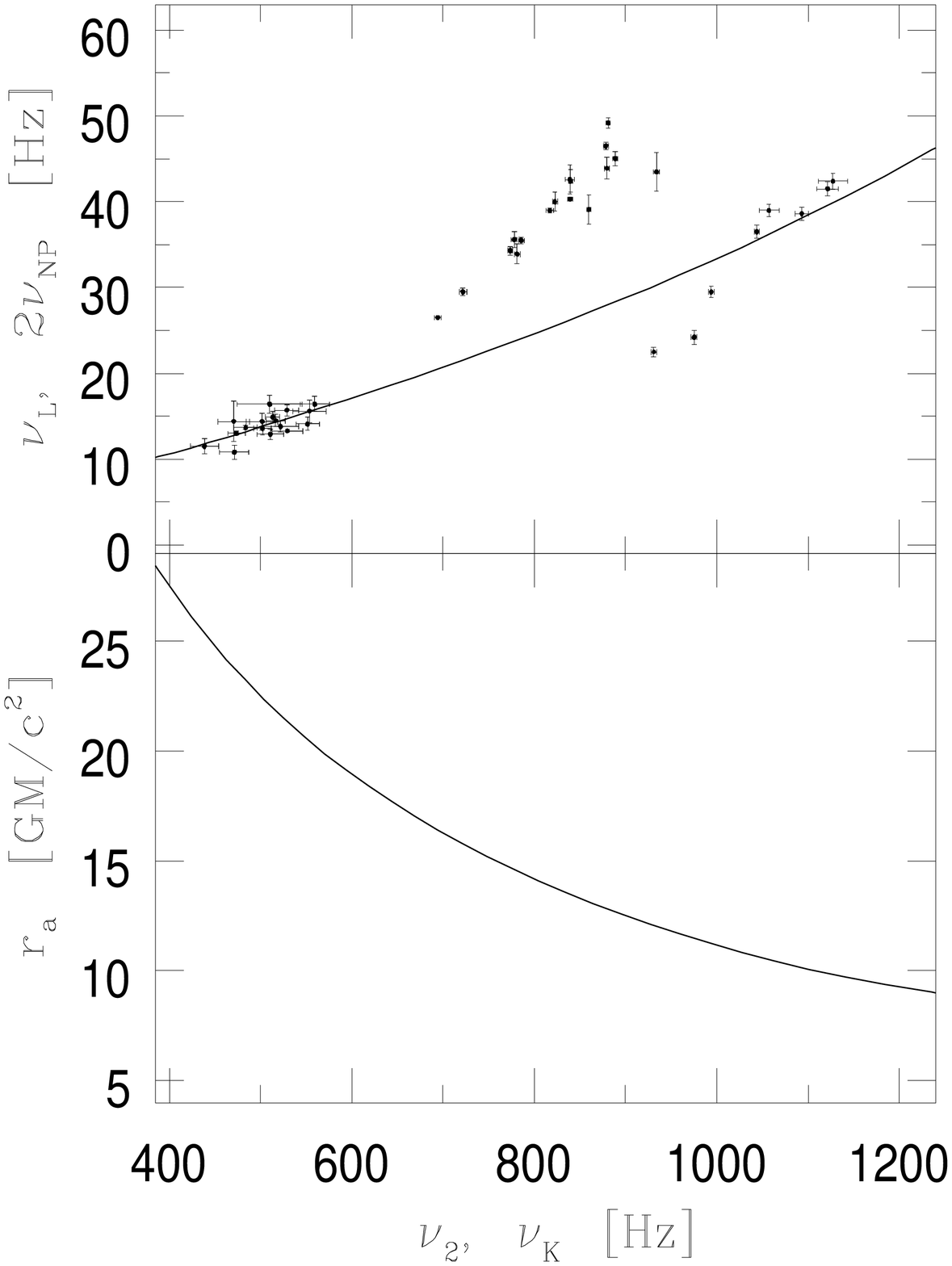}}}
\vglue-0.2 truecm
\caption {{\em Upper left}: Kilohertz QPO frequencies measured in \hbox{Sco~X-1} (data) compared with the best-fitting $\nu_{\rm AP}$ vs.\ $\nu_K$ relations given by infinitesimally eccentric geodesics around a nonrotating neutron star (solid curve), a star with $\nu_{s} = 450$~Hz constructed using the UU equation of state (dashed curve), and a black hole (dotted curve). {\em Lower left}: Circumferential radii of the geodesics used. {\em Upper right}: Comparison of the $\nu_L$ vs.\ $\nu_1$ behavior seen in \hbox{4U~1728$-$34} with the $2\nu_{\rm NP}$ vs.\ $\nu_K$ relations given by the geodesics that best fit the $\nu_1$ vs.\ $\nu_2$ data. {\em Lower right}: Apastron radii of the highly eccentric geodesics required to fit the $\nu_1$ vs.\ $\nu_2$ data. From Markovi\'c \& Lamb (2000).}\label{fig:Sco-fits}
\end{figure}

The frequencies $\nu_{\rm K}$ and $\nu_{\rm AP}$ for orbits near neutron stars are in same range as the frequencies of the kilohertz QPOs, and it was initially thought that the frequencies of the kilohertz QPOs could be fit by $\nu_{\rm AP}$--$\nu_{\rm K}$ relations constructed using slightly eccentric geodesics (Stella \& Vietri 1999; see also Karas 1999). However, the $\nu_{\rm AP}$--$\nu_{\rm K}$ relations initially used were inaccurate (see Markovi\'c \& Lamb 2000). When $\nu_{\rm AP}$ and $\nu_{\rm K}$ are computed correctly, the best-fitting frequency relations that can be constructed using infinitesimally or even moderately eccentric geodesics miss many of the measured frequency pairs by $\sim\,$30\% (see the left panel of Fig.~\ref{fig:Sco-fits}). Only by using highly eccentric geodesics with apastron radii $\sim 3$--4 times their periastron radii can the kilohertz QPO frequencies in \hbox{Sco~X-1} and the other kilohertz QPO sources be fit (Markovi\'c \& Lamb 2000). But the frequencies $\nu_{\rm NP}$ and $2\nu_{\rm NP}$ of these geodesics do not track the frequencies of the low-frequency QPOs (see the right panel of Fig.~\ref{fig:Sco-fits}).

Another issue is that numerical computations have shown that the dominant frequencies that would be generated by orbiting luminous or obscuring clumps are harmonics and sidebands of $\nu_r$, not $\nu_{\rm K}$, $\nu_{\rm AP}$, and $\nu_{\rm NP}$ or $2\nu_{\rm NP}$ (Markovi\'c \& Lamb 2000). Also, any clumps are tidally disrupted in less than 100 orbits unless they are tiny and have a mean density $\ga10^5$ the interclump density. This limits the fractional modulation that could be produced by clumps to $\ll 10^{-6}$. In the precessing clump model, the closeness of burst oscillation frequencies to ${\Delta\nu}$ or $2{\Delta\nu}$ is purely coincidental.

The precessing ring model (Psaltis \& Norman 2000) assumes there is a ``resonant ring'' of material at some special radius in the inner accretion disk and that this ring produces X-ray oscillations mainly at $\nu_{\rm K}$, $\nu_{\rm AP}$, and $2\nu_{\rm NP}$. In the inner disk, eccentric motions are quickly circularized. Hence only nearly circular motions can persist. The deviation of the best-fitting $\nu_{\rm AP}$ vs.\ $\nu_{\rm K}$ relations for approximately circular orbits from the measured kilohertz QPO frequencies by $\sim 50$--100~Hz or more is therefore an issue for this model. It has been argued (Psaltis 2000) that hydrodynamic effects may shift $\nu_r$ and $\nu_K$ by the $\sim 30$\% needed to bring the frequencies of nearly circular orbits in a ring of radial width $\sim 0.01\,r$ into agreement with the data, if the height of the ring is $\sim 0.2\,r$ and varies with radius. However, a height this large appears inconsistent with the narrow ring approximation used in the precessing ring model (Markovi\'c \& Lamb 2000). As noted in \S3.1, the physics that creates the ring and determines its radius has not been specified, so the model makes no definite predictions about the frequencies that are generated or whether they vary with the accretion rate. In this model, the closeness of burst oscillation frequencies to ${\Delta\nu}$ or $2{\Delta\nu}$ is again purely coincidental.

{\em Disk oscillation models}.---The frequencies and frequency behaviors of the $g$, $p$, and $c$-modes of relativistic accretion disks are incompatible with the observed frequencies and frequency behaviors of the kilohertz QPOs. Unlike black holes, the dimensionless angular momentum parameter $j$ (see \S3.1) of realistic neutron stars is $\la 0.2$. Hence, if the Keplerian disk flow extends inward to the ISCO, the highest-frequency $g$ and $p$-modes around 1.5--$2\,M_{\sun}$ neutron stars have frequencies $\la 600$~Hz, too low to be consistent with the frequencies of the kilohertz QPOs. Their frequencies are also insensitive to the mass accretion rate (see \S3.1), unlike the frequencies of all the QPOs seen in neutron stars. If radiation drag terminates the nearly circular Keplerian flow well outside the ISCO, these modes do not exist.

The frequencies of the fundamental $w$-modes are highest when the nearly circular Keplerian flow extends inward to the ISCO (Markovi\'c \& Lamb 1998). These highest frequencies are consistent with the highest observed frequencies of the LFQPOs ($\sim\,$80~Hz) only if these neutron stars have spin rates $\sim\,$900~Hz. If the nearly circular Keplerian flow ends well outside the ISCO, the frequencies of the $w$-modes for a given stellar spin rate will be correspondingly lower.

Although the properties of these disk modes appear inconsistent with the properties of the neutron star QPOs detected to date, it is worth searching for them in power spectra, because detecting and identifying them would provide valuable information about the structure of the inner disk and the stellar spin rate.

\subsection{Comparison of Models with Black Hole QPOs}

The frequencies of the high-frequency QPO pairs seen in \hbox{GRS~1915$+$105} and \hbox{GRO~J1655$-$40} (see Table~\ref{table.BH-HFQPOs}) appear to be relatively constant. The frequency of the single high-frequency QPO seen in \hbox{XTE~J1550$-$564} varies from 185--285~Hz (see \S2.2). The frequencies of the lower-frequency QPOs seen in these and other sources vary and are positively correlated with one another (see \S2.2 and \S3.4). These phenomenological differences suggest that different mechanisms are responsible for the different QPOs seen in the black hole sources.

{\em Sonic-point and beat-frequency mechanisms}.---In current models of disk accretion onto black holes, the radius of the transition from subsonic to supersonic radial inflow is close to the radius of the ISCO and typically varies by less than 1\%. However, these models neglect the drag force exerted by the radiation falling on the disk. Although the radiation pattern differs from the pattern in neutron star sources, dimensional estimates indicate that this drag force could affect the flow and might be strong enough to change the location of the sonic point. This possibility deserves further investigation

The magnetic fields of black holes are weak and axisymmetric and accreting gas enters the event horizon without converting its kinetic energy to radiation, unlike accreting gas that collides with the surface of a neutron star. Hence the particular beat-frequency mechanism described in \S3.1 cannot operate. It is conceivable that radiation from orbiting hot spots at one radius in the inner disk could modulate inflow from a larger radius, producing a modulation of the accretion with a frequency equal to the difference between the orbital frequencies at the two radii, but no specific models of this type have been developed.

{\em Orbital and relativistic precession models}.---An appealing idea is that the frequencies of the highest-frequency black hole QPOs are the orbital frequencies of gas orbiting near the ISCO. This hypothesis is compatible only with QPOs that have nearly constant frequencies. This interpretation is possible for the 450~Hz QPO in  \hbox{GRO~J1655$-$40} if its mass is equal to the mass estimated from spectral observations of the companion star ($6.3 \pm 0.5 M_{\sun}$; see Shahbaz et al.\ 1999; Greene, Bailyn, \& Orosz 2001) and it is rotating (Strohmayer 2001a); $j \approx 0.3$ is required for this mass. On the other hand, this interpretation is possible for the 67~Hz QPO in GRS~{1915$+$105} only if its mass is $\ga 33 M_{\sun}$ (the lowest inferred mass is obtained if the black hole is nonrotating; if it is rotating, the required mass is higher). This is about twice the measured mass of the black hole in this system, which is $14 \pm 4 M_{\sun}$ (Greiner 2001), indicating that if 67~Hz is an orbital frequency, the orbit is well outside the ISCO.

\begin{figure}[t]   
\vglue-0.25 truecm
\hbox{\hglue-0.5 truecm
{\epsfysize=2.5 truein\epsfbox{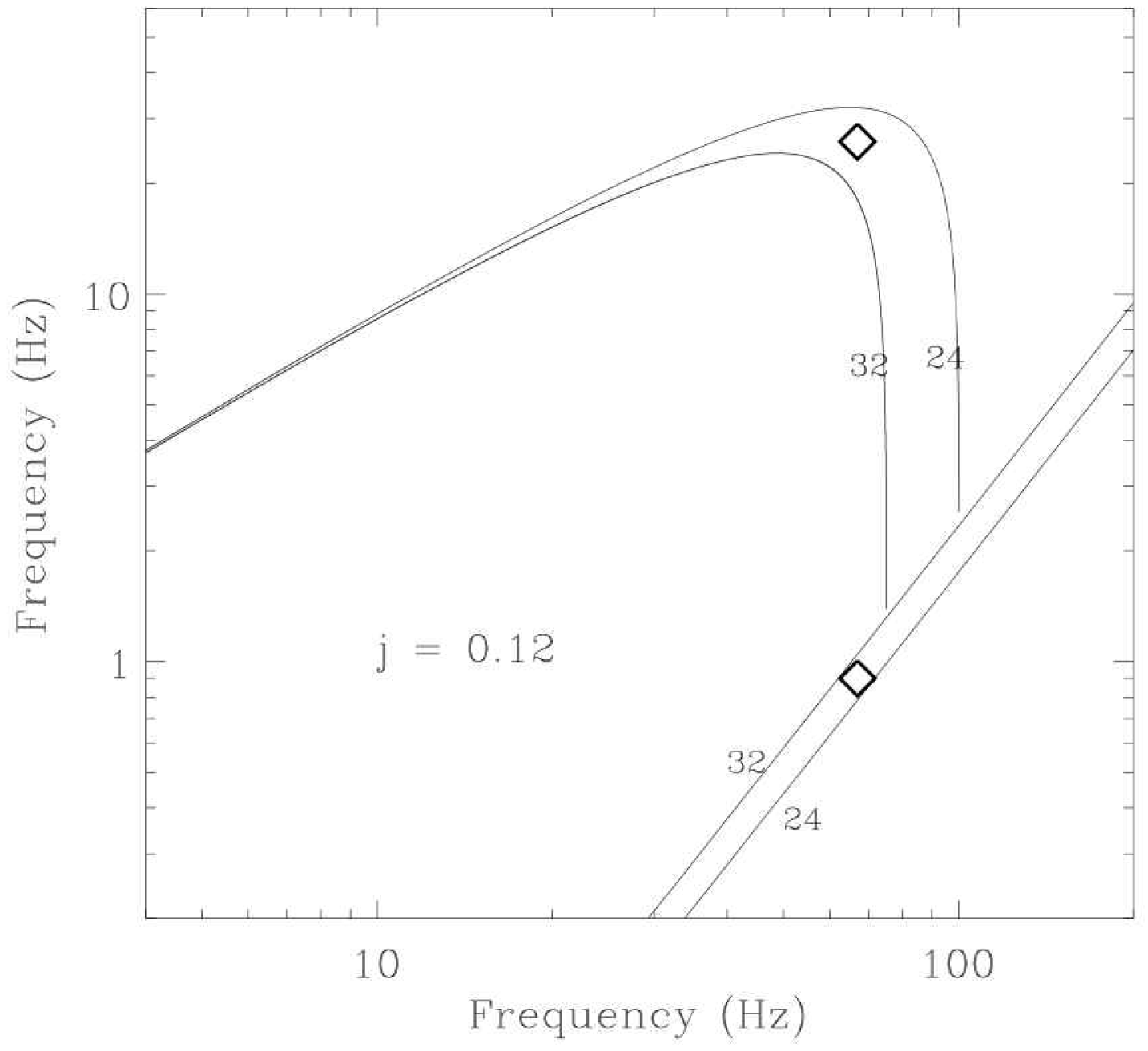}}}
\vglue-6.3 truecm
\hbox{\hglue+6.9 truecm
{\epsfysize=2.39 truein\epsfbox{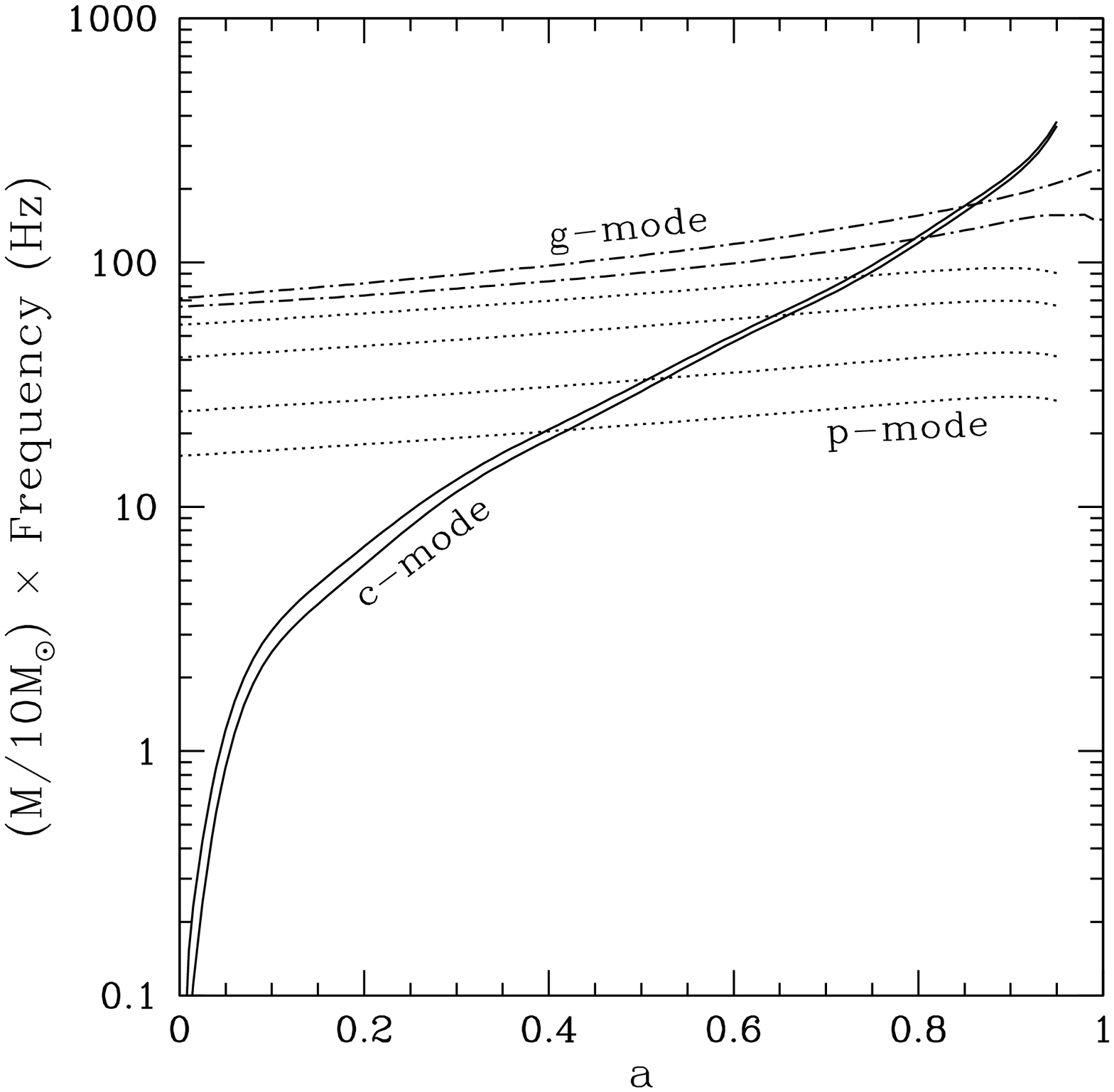}}}
\vglue-0.3 truecm
\caption {{\em Left}: The $\nu_r$ vs.\ $\nu_K$ (top) and $\nu_{\rm NP}$ vs.\ $\nu_K$ (bottom) frequency relations generated by sequences of nearly circular geodesics around $24 M_{\sun}$ and $32 M_{\sun}$ black holes with $j=0.12$. The diamonds show the proposed identifications of the difference between the 67~Hz and 40~Hz QPO frequencies in \hbox{GRS~1915$+$105} with $\nu_r$ and the 0.9~Hz noise peak with $\nu_{\rm NP}$. From Strohmayer (2001d).
{\em Right}: Dependence of the fundamental frequencies of three inner-disk oscillation modes on the spin of the black hole and, in the case of the $p$-modes, the torque on the inner edge of the disk. Here $a$ is the dimensionless angular momentum, which is denoted  by $j$ in this review. The two curves for each case are for $L/L_E = 0.01$ and 1.0. From Wagoner et al.\ (2001).}
\label{fig:BH-QPO Freqs}
\end{figure}

The dominant frequencies in the precessing clump and ring models are assumed to be the orbital, apsidal precession, and nodal precession frequencies $\nu_K$, $\nu_{\rm AP}$, and $\nu_{\rm NP}$ (see \S3.1). The frequency of the 300~Hz QPO in \hbox{GRO~J1655$-$40} could be the gravitomagnetic precession frequency at the ISCO (Cui et al.\ 1998), but only if $j \approx 1$; in this case the frequency of the 450~Hz QPO could not be the Keplerian frequency at the ISCO, which is $\ga 1$~kHz (Strohmayer 2001c). The frequencies of the 0.9~Hz, 40~Hz, and 67~Hz QPOs and peaked noise components seen in \hbox{GRS~1915$+$105} are very roughly consistent with $\nu_{\rm NP}$, $\nu_{\rm AP}$, and $\nu_K$ at the ISCO of a $\sim 30 M_{\sun}$ black hole with $j \approx 0.1$ (Strohmayer 2001d; see the left panel of Fig.~\ref{fig:BH-QPO Freqs}), but as noted above, the measured mass of this black hole is $14 \pm 4 M_{\sun}$.

The observed correlation between the frequencies of the low-frequency QPOs in some black holes is roughly consistent (within a factor of 2) with the relation between $\nu_{\rm AP}$ and $\nu_K$ for nearly circular geodesics around spinning black holes (Stella et al.\ 1999).

{\em Disk oscillation models}.---The frequencies of the $p$, $g$, and $c$-modes of the inner disk are nearly independent of the accretion rate and hence may be related to the stable, high-frequency QPOs seen in \hbox{GRO~J1655-40} and \hbox{GRS~1915$+$105} (see the right panel of Fig.~\ref{fig:BH-QPO Freqs} and Nowak \& Wagoner 1991, 1993; Perez et al.\ 1997; Nowak \& Lehr 1998; Wagoner 1999). The $p$-mode is expected to extend over only a few percent of the inner disk and the X-ray luminosity from this region is an even smaller fraction of the disk luminosity, so attention has focused on the $g$- and $c$-modes (see Wagoner et al.\ 2001). 

The 300~Hz and 450~Hz frequencies of the high-frequency QPOs in\break \hbox{GRO~J1655-40} agree with the frequencies of the $g$ and $c$ modes for $M = 5.9 \pm 1.0 M_{\sun}$ and $j = 0.917 \pm 0.024$ (Wagoner et al.\ 2001), consistent with the $6.3 \pm 0.5 M_{\sun}$ mass estimate from spectral observations of the companion star (Shahbaz et al.\ 1999; Greene, Bailyn, \& Orosz 2001). The frequency of the fundamental $w$-mode is essentially the same as that of the fundamental $c$-mode (see \S3.1), so the 450~Hz oscillation might also be generated by this mode. The high inclination of \hbox{GRO~J1655-40} makes it more likely that these nonaxisymmetric modes can produce X-ray oscillations with significant amplitudes. 

Identification of the 67~Hz high-frequency QPO in \hbox{GRS~1915$+$105} with the $g$-mode gives $M = 18.2 \pm 3.1$, consistent with the $14 \pm 4 M_{\sun}$ estimated mass of this black hole (Greiner 2001), if $j = 0.701 \pm 0.043$ (Wagoner et al.\ 2001). For such a black hole, the frequency of the $c$-mode is about 50~Hz, somewhat higher than the 40~Hz frequency of the lower high-frequency QPO in \hbox{GRS~1915$+$105}.

\subsection{Discussion}

A large number of QPOs have now been seen in accreting neutron stars and black holes and a variety of mechanisms have been suggested to explain them. An interesting question is whether the mechanisms that produce the QPOs in black holes and neutron stars are similar or different.

{\em QPO mechanisms in neutron stars and black holes}.---Different QPO mechanisms are possible in neutron stars and black holes, because of the physical differences between them. Moreover, even the same mechanism may operate differently because of these physical differences.

The existence of a stellar surface and of magnetic fields tied to the surface are probably the physically most important properties of neutron stars not shared by black holes. These properties of neutron stars are likely to affect the frequencies, waveforms, and other properties of QPOs that originate within a few stellar radii, causing differences from those originating near black holes, even if the mechanism is basically the same. Depending on their masses and the equation of state of neutron star matter, some neutron stars may have radii large enough that there is no ISCO. In contrast, there is always an ISCO around a black hole. The spectra and angular patterns of the radiation coming from the surfaces of neutron stars and the inner disks around them are expected to be significantly different from those of the radiation coming from the inner disks around black holes. As discussed in \S3.1, the drag force on accreting gas produced by the radiation coming from the surface of a neutron star is dynamically important. The neutron stars in LMXBs are thought to be the progenitors of the recycled radio pulsars (see Ghosh \& Lamb 1992; Kulkarni \& Phinney 1994; Lorimer 2001) and to have $\sim\,$10$^8$--10$^9$~G dipole magnetic fields. There is also a fairly large body of indirect evidence for such fields (see \S3.2). Surface fields $\ga 10^6$~G have dynamical effects on the accretion flow near the stellar surface (Ghosh \& Lamb 1992) and would be expected to affect the properties of QPOs that originate within a few stellar radii.

The exterior spacetimes of neutron stars and black holes are also different and hence the frequencies of  dynamical motions near them are different (see, e.g., Miller et al.\ 1998a). The two most important differences are their masses and angular momenta. For fixed $j$, the dynamical frequencies near neutron stars and black holes scale as $1/M$. The masses of black holes in LMXBs are typically $\sim 5$--10 times larger than the masses of neutron stars, so dynamical frequencies near them should be $\sim 5$--10 times lower for the same $j$. However, $j$ is limited to $\la0.2$ for realistic models of neutron stars but can approach unity for black holes. Gravitomagnetic frequencies, and hence the frequencies of the $c$- and $w$-modes, can therefore be much higher near black holes.

Because of these important differences between neutron stars and black holes, it would be surprising if the highest-frequency QPOs produced by these objects, which presumably are generated close to their surfaces and event horizons, respectively, have identical properties.

\begin{figure}   
\vglue-0.2 truecm
\hbox{\hglue+0.2 truecm
{\epsfysize=2.65truein\epsfbox{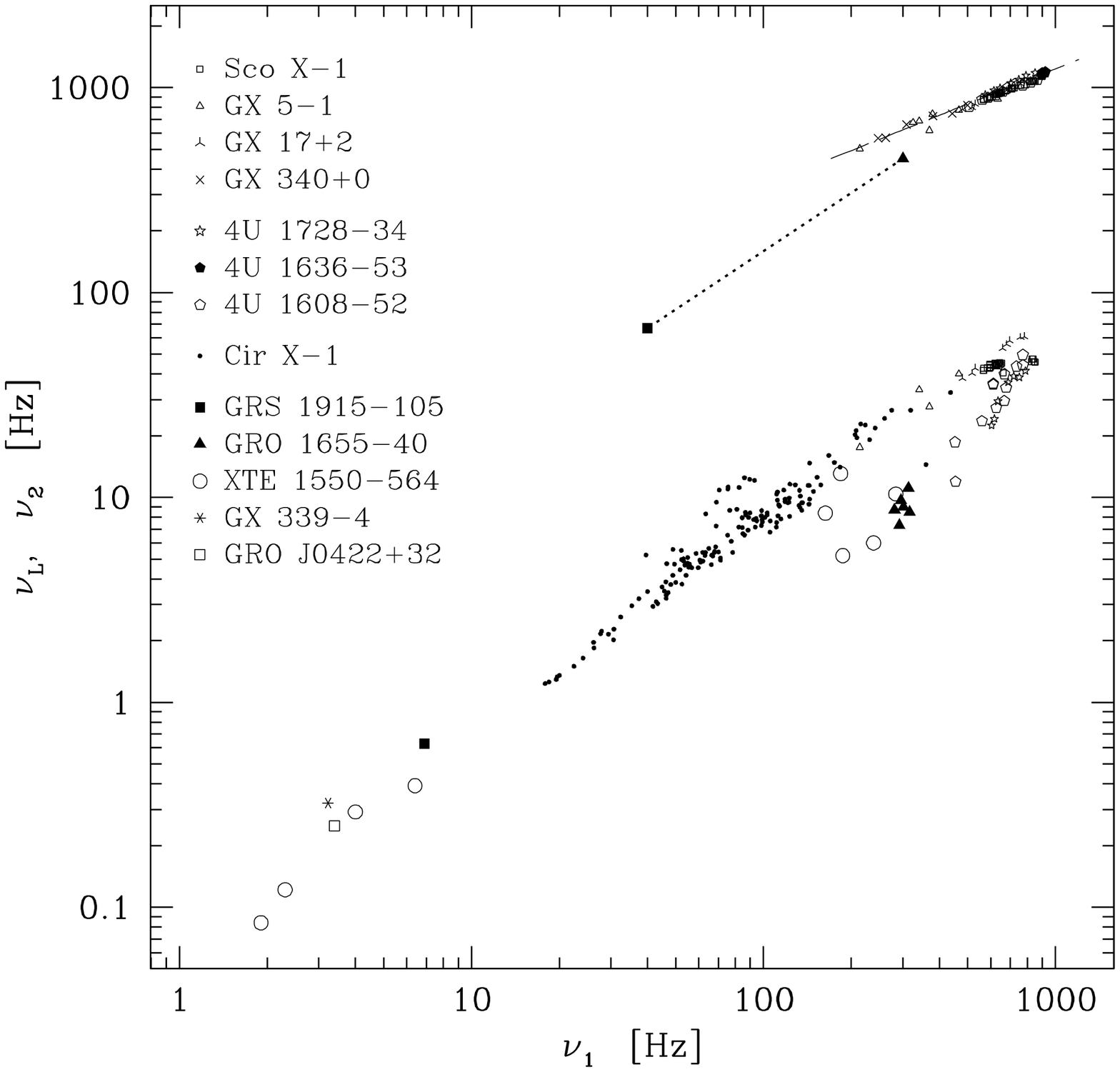}}
\hglue-0.7 truecm
{\epsfysize=2.65truein\epsfbox{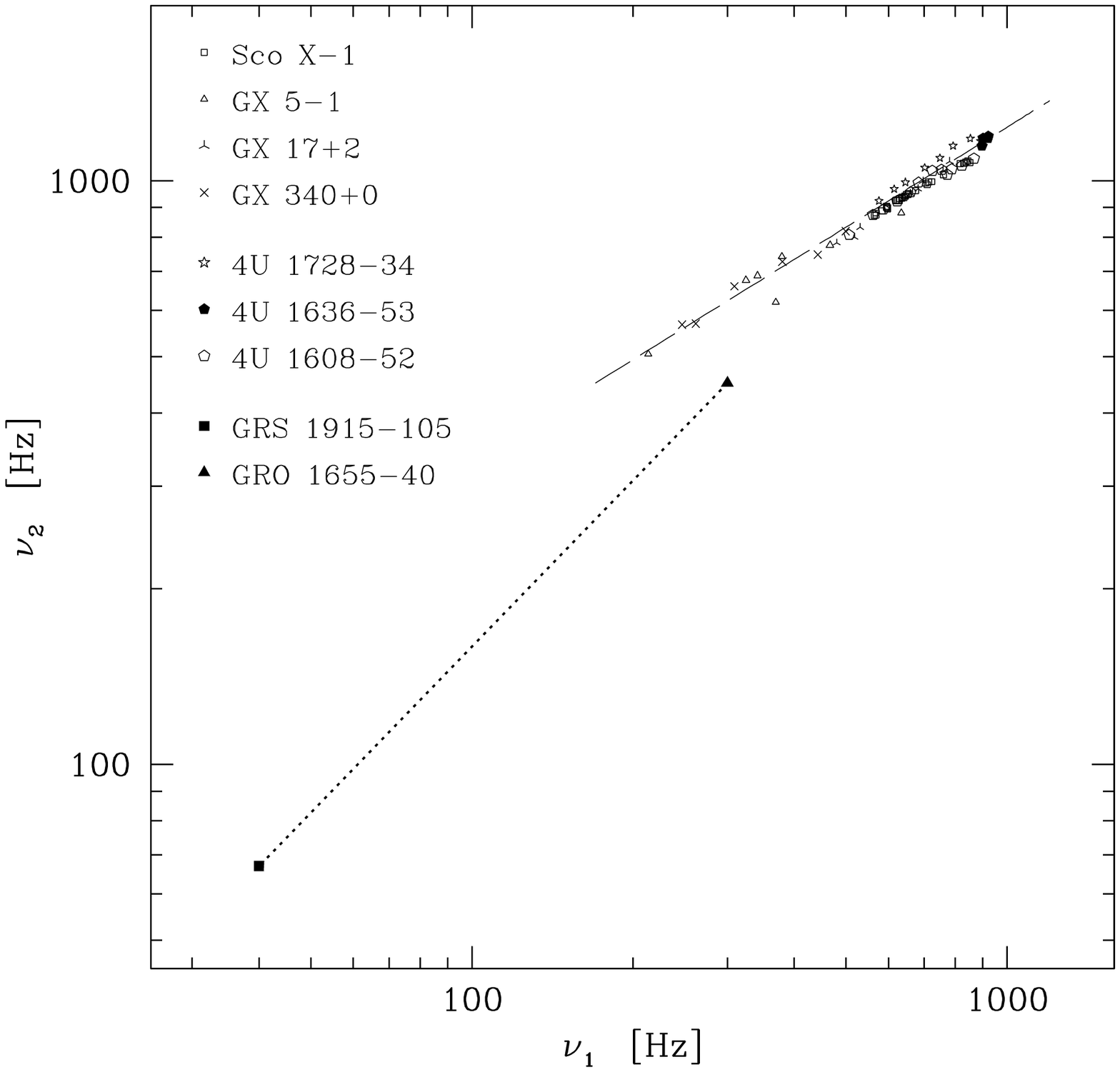}}}
\vglue-0.5 truecm
\caption{Frequencies of some of the high-, medium-, and low-frequency QPOs and peaked noise components seen in the power spectra of the LMXBs listed, plotted against the frequencies of the medium-frequency QPOs seen in these same sources. The dotted line connects the high-frequency QPO pairs seen in \hbox{GRO~J1655$-$40} and \hbox{GRS~1915$+$105}. {\em Left\/}: Frequencies in the range 0.1~Hz to 10000~Hz. {\em Right\/}: Expanded view of the 40--1000~Hz range. After Psaltis et al.\ (1999a).}
\label{fig:imbroglio}
\end{figure}

{\em Frequency correlations}.---Psaltis et al.\ (1999a; see also van der Klis 2000) have called attention to correlations between the frequencies of some of the low- and medium-frequency QPOs and noise components observed in several black hole sources and the similarity of the slopes of these correlations to the slopes of some of the frequency correlations observed in neutron star sources (see Fig.~\ref{fig:imbroglio}). These frequency variations are thought to reflect changes in the accretion rate. Psaltis et al.\ have suggested that some of these power spectral features may have similar origins in neutron stars and black holes.

Their high and relatively stable frequencies and other properties seem to set the high-frequency QPO pairs seen in \hbox{GRO~J1655$-$40} and \hbox{GRS~1915$+$105} apart from the other black hole QPOs (see Fig.~\ref{fig:imbroglio} and \S2.2). As discussed \S3.3, the highest frequencies may be orbital frequencies or the frequencies of disk oscillations near the ISCO. It appears that the differences between the highest frequencies seen in these two sources are not due solely to their different masses. If so, the mechanisms involved must be at least somewhat different.

The frequencies of the upper kilohertz QPOs also stand apart in  Figure~\ref{fig:imbroglio}. These are thought to be the orbital frequency at the inner edge of the nearly circular flow around the neutron star (see \S3.2). If the QPO mechanisms in neutron stars and black holes are indeed the same, it is somewhat puzzling that this fundamental frequency has not been detected in most black holes.

The frequency correlations of the lower kilohertz and low-frequency QPOs in atoll sources such as \hbox{4U~1728$-$34} and \hbox{4U~1608$-$52} are separated from the main correlation line established by the frequency correlations of the lower kilohertz and horizontal-branch QPOs in the Z sources and the peculiar source \hbox{Cir~X-1} by a factor $\sim\,$2 and seem to have a much steeper slope. The frequencies of the medium- and low-frequency QPOs in \hbox{XTE~J1550$-$564} and \hbox{GRO~J1655$-$40} are separated from the main correlation line by a factor $\sim\,$2 and do not establish a clear correlation.

These examples indicate that not all QPO frequencies are correlated in the same way and hence that it is possible that different mechanisms are involved in different types of sources. The main correlation line that is evident in Figure~\ref{fig:imbroglio} is established by correlations of the lower kilohertz and low-frequency QPOs in the Z sources and especially in the peculiar source \hbox{Cir~X-1} and the correlations of the two low-frequency QPOs in several black holes, which fall approximately on the same line in the frequency-frequency plane. It is also worth noting that it is the frequencies of the two most prominent band-limited noise components in the power spectrum of \hbox{Cyg~X-1}, not the QPO frequencies, that appear to follow roughly the main correlation seen in Figure~\ref{fig:imbroglio} (see Nowak 2001).

{\em Physical origins of frequency correlations}.---There are several possible ways that the frequency correlations evident in Figure~\ref{fig:imbroglio} might be explained. One possibility is that two QPOs are generated at a single radius and scale with radius in approximately the same way, so that when the radius changes with accretion rate, they trace out a one-parameter curve in the frequency-frequency plane. For example, at large distances from neutron stars and black holes ($r \ga 10\, GM/c^2$), the nodal and apsidal precession frequencies satisfy $\nu_{\rm NP} \approx j\pi^{1/5}(2/3)^{6/5} (\nu_{\rm PA})^{6/5}$ (see \S3.1 and Markovi\'c 2000) and are therefore almost proportional to one another. If $\log \nu_{\rm NP}$ is plotted versus $\log \nu_{\rm AP}$, the points will fall on a line with a slope of 1.2, close to the slope of the main correlation in Figure~\ref{fig:imbroglio}.  Moreover, both $\nu_{\rm NP}$ and $\nu_{\rm PA}$ are steep functions of radius (in relativistic units, $2\pi{\hat\nu}_{\rm PA} \approx 3{\,\hat r}^{-5/2}$ while $2\pi{\hat\nu}_{\rm NP} \approx 2j{\hat r}^{-3}$), so only a modest variation in the radius at which the QPOs are produced can produce a large variation in their frequencies. Such a mechanism might be relevant to the frequencies of the low frequency and lower-kilohertz QPOs generated by \hbox{Cir~X-1}, which vary by a factor $\sim\,$20. If either of these are orbital frequencies, the radius at which they are generated would have to vary by a factor $\sim\,7$ or more, which would be difficult to understand (see below), whereas if they are precession frequencies, a variation in radius of a factor of 1.8 would be sufficient. The \hbox{Cir~X-1} QPO frequencies are in approximately the right range to be precession frequencies (see Stella et al.\ 1999) and a slope of 1.2 may describe their correlation better than a slope of 1.

Sometimes an apparent power-law frequency correlation over a limited range of frequencies may be explained by a model that does not predict a power-law relation. For example, the phenomenological power-law relation $\nu_{\rm HBO} \propto \nu_s \nu_2^{1.6}$ accurately describes the observed correlation between the frequencies of the HBO and upper kilohertz QPO in the Z sources, but so does the relation $\nu_{\rm HBO} = \nu_s (K\nu_2^{0.16} - 1)$, which is predicted by the magnetospheric beat-frequency model (see Psaltis et al.\ 1998b).

These considerations underscore the importance of paying attention to the whole spectrum of information we now have on the phenomenology of the different QPOs, not just their frequencies. Of particular importance is how the individual QPO frequencies vary with accretion rate, because this variation is an important guide to the physical mechanism(s) involved. 

{\em The challenge of variable frequencies}.---I have already noted that large variations in the frequencies of QPOs are challenging for physical theories. This is particularly true if the oscillations are highly coherent; understanding large variations in the frequencies of peaked or band-limited noise components is generally less demanding. Several QPO models are based on the existence of one or more special radii in the disk that decrease substantially as the accretion rate increases. A key question is, what physics picks out a special radius and causes it to vary by a substantial factor as the accretion rate changes by, e.g., $\sim\,$50\%?

To my knowledge, the only mechanism so far identified that picks out a special radius that decreases markedly as the accretion rate increases is the radiation drag mechanism  that is at the heart of the  sonic-point model (see \S3.1). In its original form, this mechanism can explain a variation in the radius of the sonic point by a factor $\sim\,$3, from $\sim R_{\rm star}$ to $\sim 3R_{\rm star}$. The radius and radius range could be greater if radiation drag only needs to act on a small fraction of the accretion flow, such as gas on the surface of the disk, to make a QPO mechanism work. It is possible that the radiation-drag mechanism could also work in black hole systems, although this possibility has not been studied in any detail.

The radiation drag mechanism has been invoked to explain the existence of the special radii needed in models other than the sonic-point model. It is important to bear in mind that the radiation force has important dynamical effects on the accretion flow over an extended region, where it changes both the orbital and the radial velocities of the flow. These effects, as well as the existence of a sonic radius, must be taken into account in any self-consistent model of the formation of QPOs near neutron stars.

\section{Using QPOs to Study Dense Matter and Strong-Field Gravity}
\label{sec:Using-QPOs}

As mentioned in the Introduction, almost all QPO models developed since the launch of {\em RXTE\/} have considered the effects of strong-field gravity. In this sense, the QPOs themselves provide information on gas dynamics and radiation transport in strongly curved spacetimes. In this section I discuss briefly how the kilohertz QPOs can be used to extract further information about the properties of strong  gravitational fields and dense matter.

\begin{figure}[t]   
\vglue-0.9 truecm
\hbox{\hglue+0.5 truecm
{\epsfysize=2.75truein\epsfbox{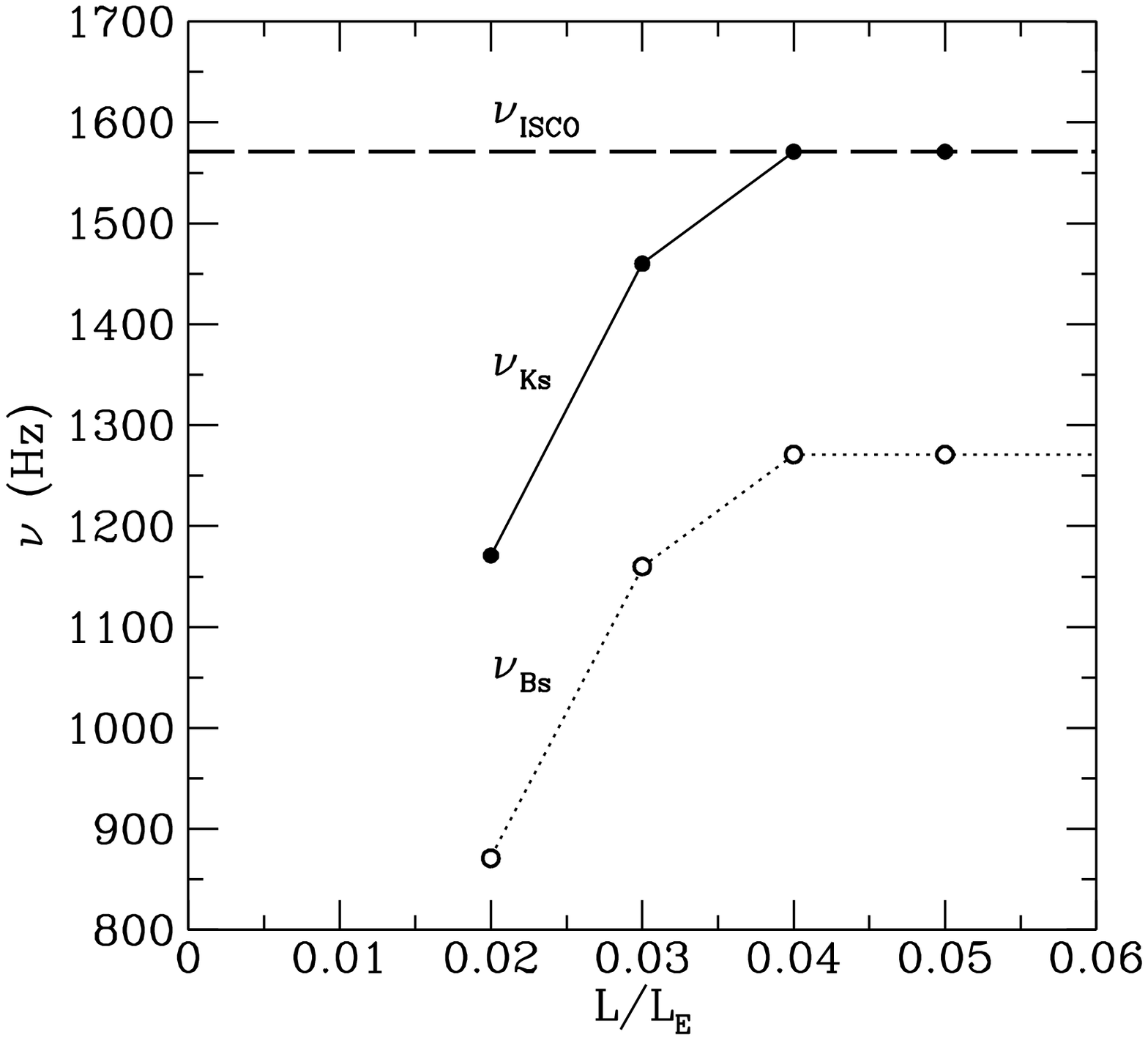}}
\hglue-1.3 truecm
{\epsfysize=2.75truein\epsfbox{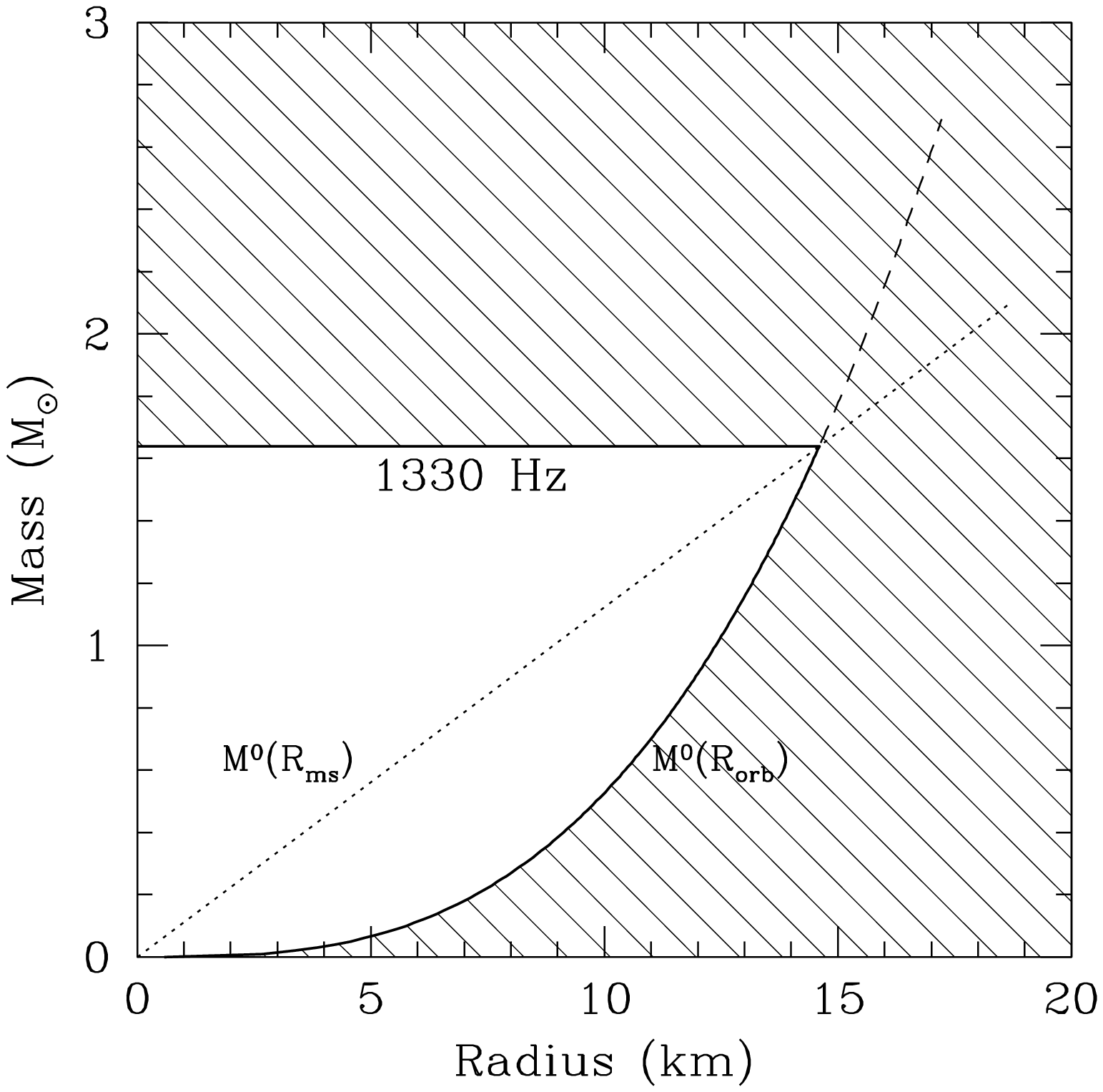}}}
\vglue-0.7 truecm
\caption{{\em Left\/}: Example of the dependence of the upper and lower kilohertz QPO frequencies on the accretion luminosity found in calculations using the sonic-point beat-frequency model. Once the sonic point reaches the ISCO, the curve flattens and hence the plateau frequency is the orbital frequency at the ISCO.
  {\em Right\/}: Mass-radius plane, showing the region allowed for a nonrotating neutron star with $\nu_2^\ast = 1330$~Hz (see text). The star's representative point must lie in the unhatched region enclosed by the solid line.
  After Miller et al.\ (1998a). }
  \label{fig:pie-plateau}
\end{figure}

Assuming that the upper kilohertz QPO at $\nu_2$ is produced by orbital motion of gas near the neutron star, its behavior can be used to investigate orbits in the region of strongly curved spacetime near the star. For example, it may be possible to establish the existence of an innermost stable circular orbit (ISCO) around some neutron stars in LMXBs (see Kaaret \& Ford 1997; Miller et al.\ 1998b, 1998c; Lamb et al.\ 1998b). This would be an important step forward in our understanding of strong-field gravity and the properties of dense matter, because it would be the first confirmation of a prediction of general relativity in the strong-field regime.

\begin{figure}[t]   
\vglue-0 truecm
\hbox{\hglue 0.4 truecm
{\epsfysize=5 truecm\epsfbox{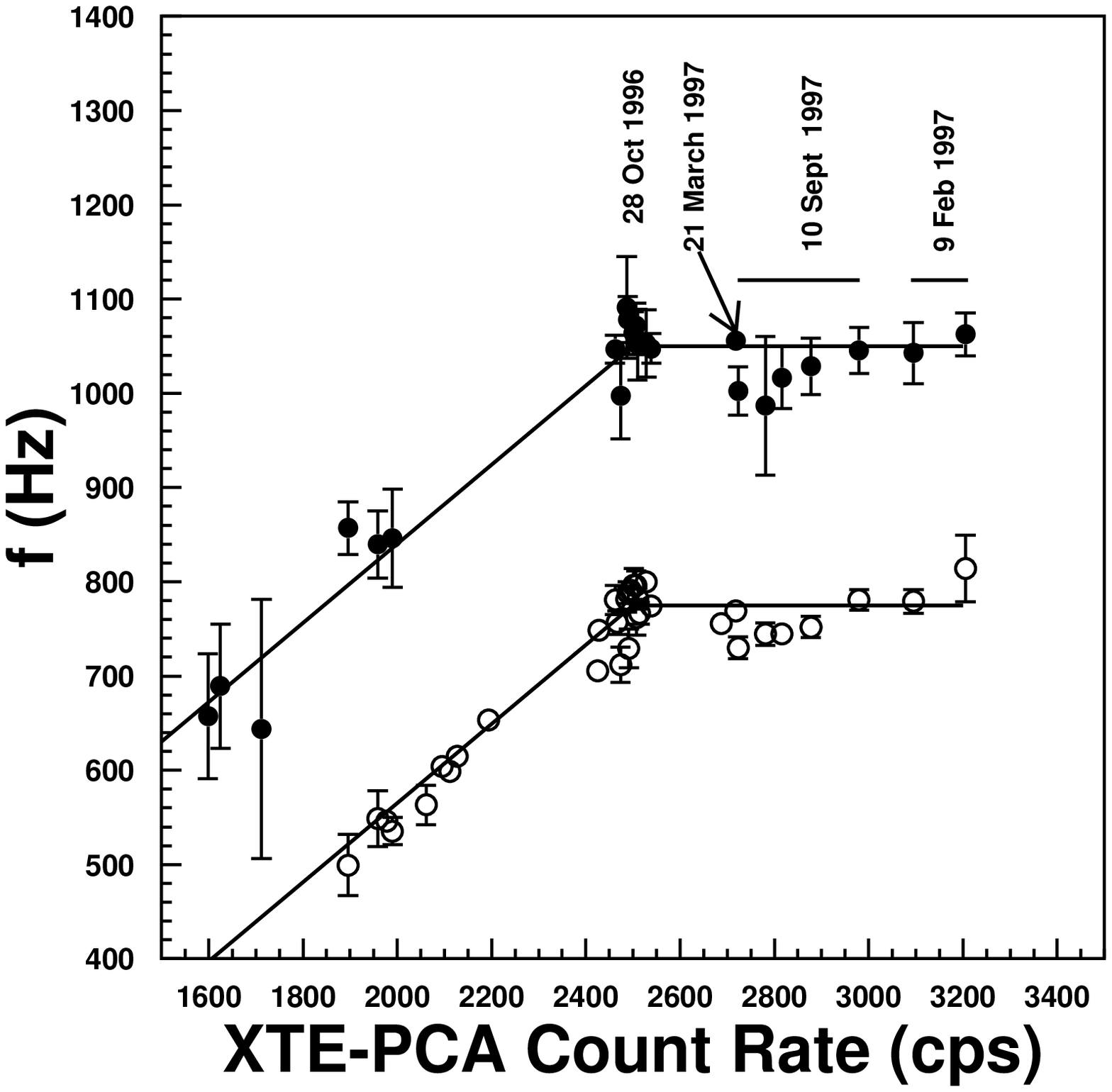}}}
\vglue-5.4 truecm
\hbox{\hglue+5.75 truecm
{\epsfysize=5.68 truecm\epsfbox{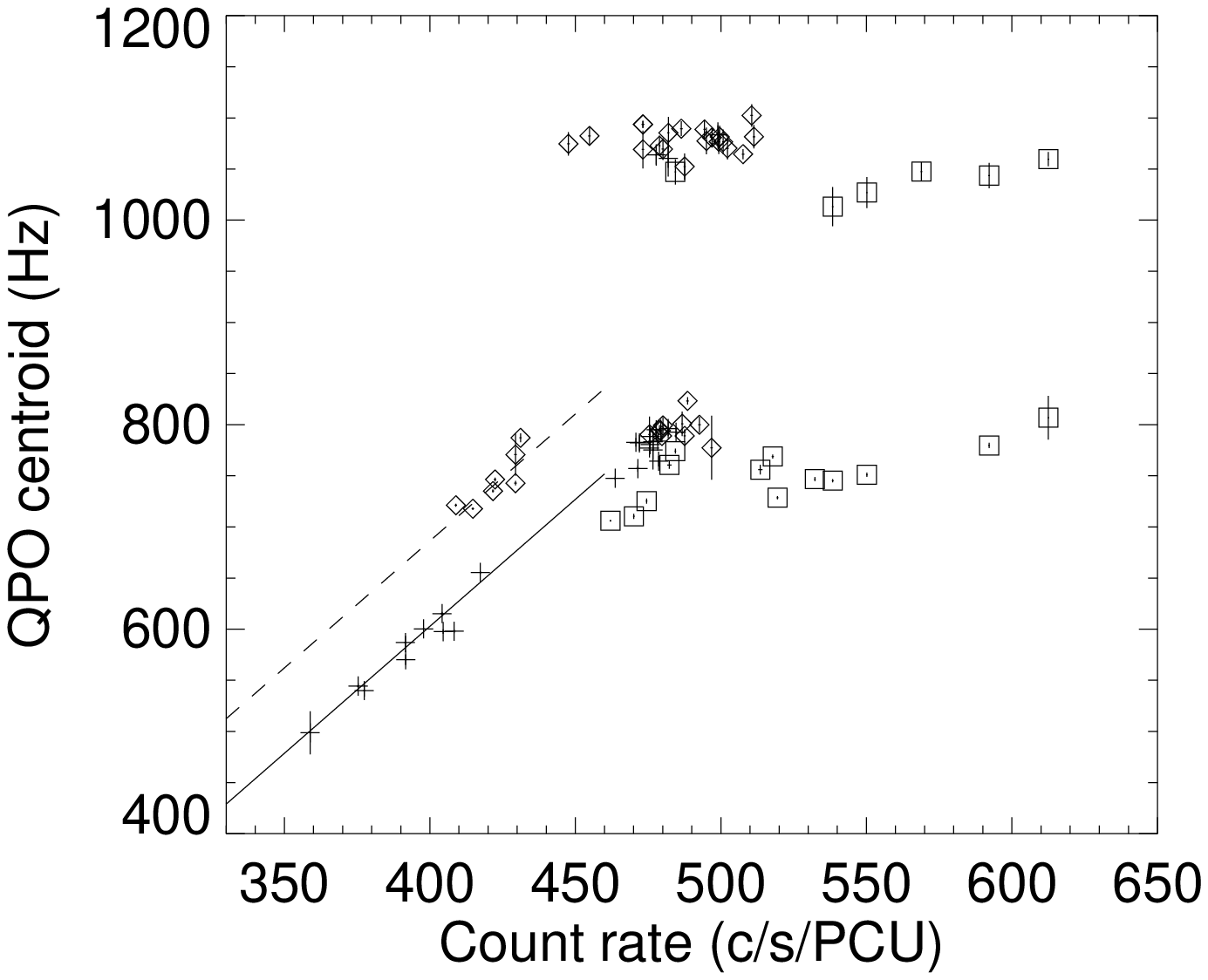}}}
\vglue-0.4 truecm
\caption{Kilohertz QPO frequency vs.\ countrate tracks in \hbox{4U~1820$-$30} observed by Zhang et al.\ (1998) (left) and Kaaret et al.\ (1999) (right).}
\label{fig:1820-tracks}
 \end{figure}

Possible signatures of the ISCO have been discussed by Miller et al.\ (1998a) and Lamb et al.\ (1998a, 1998c). The sonic-point model predicts that the frequencies of both kilohertz QPOs will increase with the accretion luminosity until the sonic radius---which moves inward as the mass flux through the inner disk increases---reaches the ISCO, at which point it will become independent of the accretion luminosity (see the left panel of Fig.~\ref{fig:pie-plateau}). The plateau frequency is the orbital frequency of the ISCO. Zhang et al.\ (1998), Kaaret et al.\ (1999), and Bloser et al.\ (1999) have reported seeing behavior similar to this in \hbox{4U~1820$-$30} (see Fig.~\ref{fig:1820-tracks}).

Although the observed variations of $\nu_1$ and $\nu_2$ with countrate (and X-ray flux) are very similar to the relation predicted by the sonic-point model, several questions must be answered before these results can be accepted as compelling evidence of an ISCO. For example, how robust is the predicted relation between QPO frequency and ${\dot M}$? What is the expected behavior of the kilohertz QPO amplitudes as the sonic point reaches the radius of the ISCO? Naively, one would expect the amplitude of the QPO at the beat frequency to decrease. Could the apparent plateau be an artifact of shifts in the $\nu_1$ and $\nu_2$ tracks like those seen in other sources (see Fig.~\ref{fig:1608-tracks})?

\begin{figure}[t]   
\vglue-0.9 truecm
\hbox{\hglue+0.3 truecm
{\epsfysize=2.75truein\epsfbox{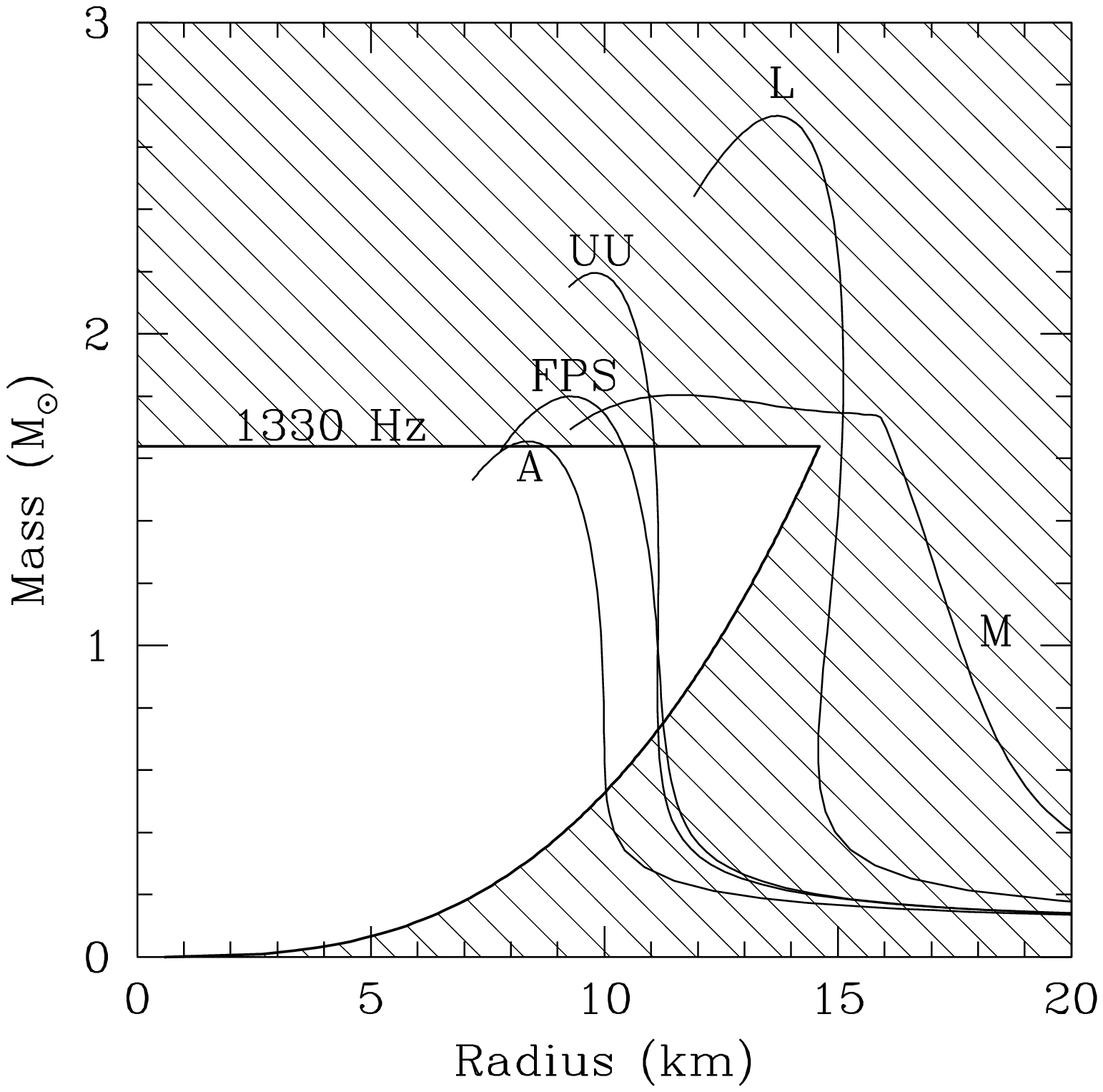}}
\hglue-1.0 truecm
{\epsfysize=2.75truein\epsfbox{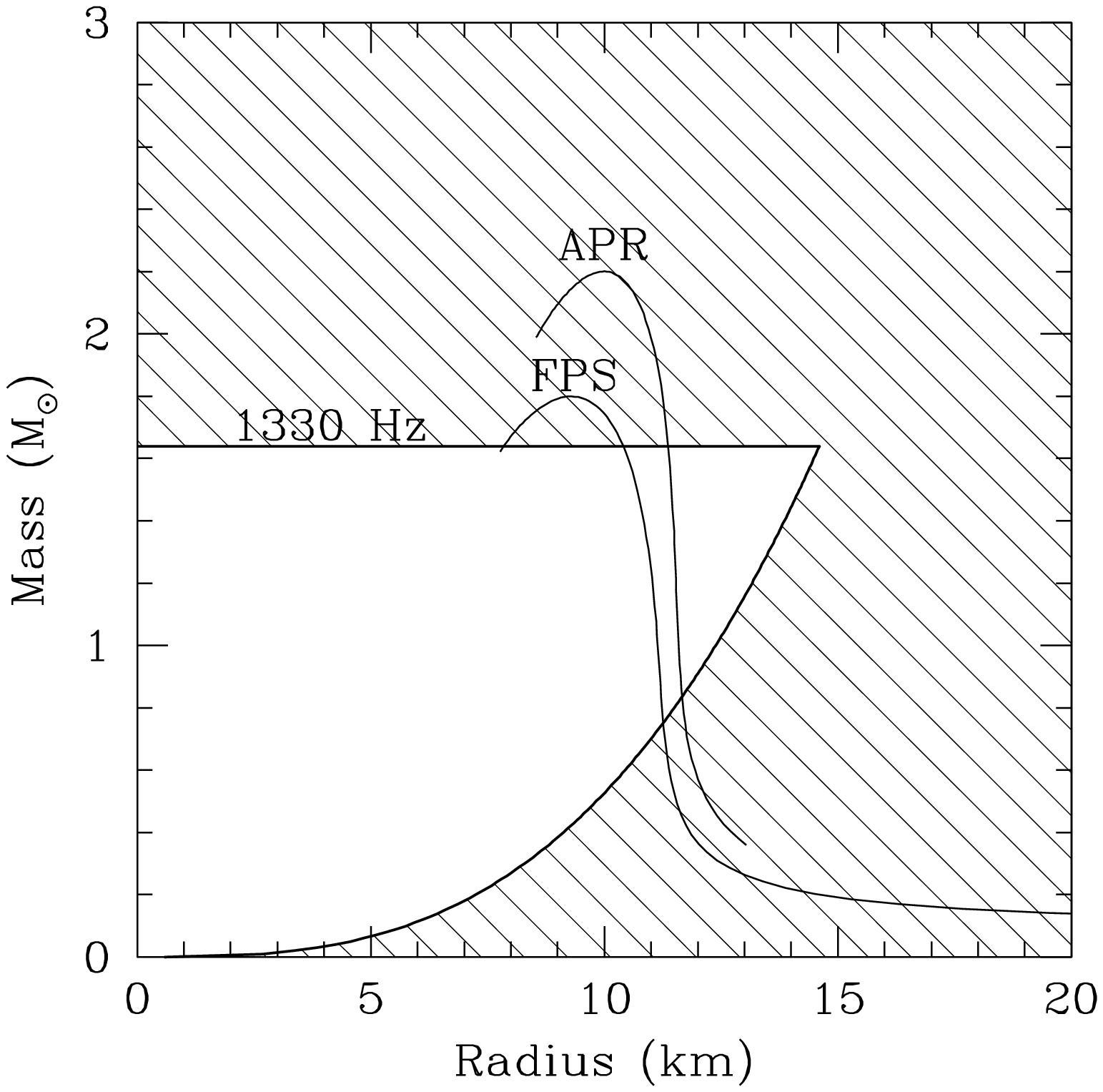}}}
\vglue-0.5 truecm
\caption{The region of the $M$-$R$ plane allowed for a nonrotating star in which a 1330~Hz orbital frequency has been observed, compared with the mass-radius relations given by five representative neutron star equations of state (left) and two recent realistic equations of state (right). }
  \label{fig:EOS-constraints}
\end{figure}

{\em Using the kilohertz QPOs to determine the masses, radii, and spin rates of neutron stars}.---Assuming only that the frequency of a given kilohertz QPO is an orbital frequency, one can extract constraints on the mass and radius of the neutron star (Miller et al.\ 19998; Lamb et al.\ 1998). These constraints can be determined for any stellar spin frequency (see Miller et al.\ 1998b; Miller, Lamb, \& Cook 1998), but it is easier to see how they are obtained for a nonrotating star. I therefore consider this case here.

The right panel of Figure~\ref{fig:pie-plateau} shows how to construct constraints on the mass and radius of a nonrotating neutron star, given $\nu_2^\ast$, the highest orbital frequency observed in the source. $R_{\rm orb}$ must be greater than the stellar radius, so the star's representative point must lie to the left of the (dashed) cubic curve $M^0(R_{\rm orb})$ that relates the star's mass to the radius of orbits with frequency $\nu_2^\ast$. The high coherence of the oscillations constrains $R_{\rm orb}$ to be greater than $R_{\rm ms}$, the radius of the innermost stable orbit, which means that the radius of the actual orbit must lie on the $M^0(R_{\rm orb})$ curve below its intersection with the (dotted) straight line $M^0(R_{\rm ms})$ that relates the star's mass to $R_{\rm ms}$. These requirements constrain the star's representative point to lie in the unhatched, pie-slice shaped region enclosed by the solid line. The allowed region shown is for $\nu_2^\ast = 1330$~Hz, the highest value of $\nu_2$ observed in \hbox{4U~0614$+$09} (van Straaten 2000), which is also the highest value so far observed in any source.

The left panel of Figure~\ref{fig:EOS-constraints} shows how this allowed region compares with the mass-radius relations given by five representative equations of state (for a description of these EOS and references to the literature, see Miller et al.\ 1998a). If \hbox{4U~0614$+$09} were not rotating, EOS~L and M would both be excluded. However, \hbox{4U~0614$+$09} is undoubtedly rotating and probably has a spin frequency $\ga350$~Hz, in which case EOS~M may be allowed. However, EOS~L is excluded for any spin rate. The right panel of Figure~\ref{fig:EOS-constraints} shows how the allowed region compares with the mass-radius relations given by the FPS EOS and the most recent realistic EOS, denoted APR (Akmal, Pandharipande, \& Ravenhall 1998). Both are allowed for nonrotating neutron stars, but the FPS EOS may be excluded if \hbox{4U~0614$+$09} is rapidly rotating.

If one can establish that a given QPO frequency is the frequency of the ISCO, for example by detecting a stable plateau in the QPO frequency-luminosity relation, the mass of the star can be determined. The right panel of Figure~\ref{fig:pie-plateau} shows how this can be done for a nonrotating star. If an observed value of $\nu_2$ is the frequency of the ISCO, the radius of the orbit involved must lie on the diagonal dashed line as well as on the cubic curve that corresponds to $\nu_2$, i.e., it is at the intersection of the two. This fixes the mass of the star for a given proposed EOS (Miller et al.\ 1998a, 1998b, 1998d; Lamb et al.\ 1998a, 1998c). As a specific example, if \hbox{4U~1820$-$30} has a spin rate of 290~Hz and the 1060~Hz plateau frequency observed in this source is the frequency of the innermost stable orbit, general relativistic calculations show that the neutron star must have a mass close to $2.3\,M_{\sun}$ for the few equations of state that give a solution. If confirmed, such a high mass would imply that the three-nucleon interaction is highly repulsive at high densities and probably that there is no quark matter at the center of even such a massive neutron star. These conclusions would be of fundamental importance for nuclear physics as well as for astrophysics.

Once they have been firmly established, specific QPO models will make it possible to derive further constraints on the properties of neutron stars. For example, the fits shown in Figure~\ref{fig:SPBFComparison} of the sonic-point beat-frequency model to the kilohertz QPO behavior observed in \hbox{Sco~X-1}, \hbox{4U~1608$-$52}, \hbox{4U~1728$-$34}, and \hbox{4U~1820$-$30} give the masses and spin rates listed in Table~\ref{table.NSProperties}. The effect of the star's angular momentum on its structure and the spacetime were neglected, which is why the inferred mass of \hbox{4U~1820$-$30} is $2.0\,M_{\sun}$, rather than $2.3\,M_{\sun}$.

\begin{table}[b]
\begin{center}
\begin{minipage}{102mm}
\caption{Inferred Properties of Neutron Stars}
\label{table.NSProperties}
\vspace{3pt}
\begin{tabular}{cccc}
\tableline
\tableline
\noalign{\kern 4pt}
Source   &\ \ Mass ($M/M_{\sun}$)   &\ \ \ Spin Rate      &\ \ \ $\chi^2/{\rm dof}$  \\
\noalign{\kern 4pt}
\tableline
\noalign{\kern 3pt}
\hbox{Sco~X-1}             &1.59        &352           &95.5/46\\
\noalign{\kern 1pt}
\hbox{4U~1608$-$52}    &1.98        &$\;\;\;\;\,309.5^a$     &20.6/10\\
\noalign{\kern 1pt}
\hbox{4U~1728$-$34}    &1.74        &364           &8.5/6\\
\noalign{\kern 1pt}
\hbox{4U~1820$-$30}    &2.0$^b$     &279           &7.0/18\\
\noalign{\kern 3pt}
\tableline
\end{tabular}
$^a$Set equal to one-half the burst oscillation frequency. $^b$Set equal to the mass given by equating the frequency of the plateau to the frequency of the ISCO.
\end{minipage}
\end{center}
\end{table}

\section{Concluding Remarks}
\label{sec:conclusions}

The launch of the {\em Rossi X-Ray Timing Explorer\/} has brought a flood of new information about neutron stars and black holes in LMXBs, stimulating imaginative new thinking about these systems. A variety of mechanisms have been proposed to explain the many QPOs that have been discovered. We are only beginning to explore and test them. It is likely that some mechanisms which generate QPOs have not yet been identified. New observational results that will help advance our understanding are being reported almost daily. 

Despite uncertainty about the precise physical mechanisms responsible for generating many of the QPOs seen in neutron star and black hole systems, there is good evidence that the upper kilohertz QPO in neutron stars is produced by orbital motion of gas in the strong gravitational field near the star. Making only this minimal assumption, important new constraints on the masses and radii of the neutron stars in LMXBs can be derived. There are indications that the signature of an innermost stable orbit has been seen in \hbox{4U~1820$-$30}. If this is confirmed, it will be the first detection of a strong-field general relativistic effect.

The ability to study variations in the emission of neutron stars and black holes on the dynamical time scales near them has made it possible to address questions of fundamental physics. It is important to continue this effort.

\acknowledgments

I thank Y. Chen, P. Jonker, M. van der Klis, D. Markovi\'c, M. M\'endez, M. Nowak, D. Psaltis, R. Remillard, J. Swank, W. Zhang for many valuable discussions. I am also grateful to Y. Chen M. van der Klis, and D. Markovi\'c for their comments on a draft of this review. This research was supported in part by NASA grant NAG~5-8424 and NSF grant AST~0096399.

\end{document}